\documentclass{emulateapj}

\def\eqq#1{Equation~(\ref{#1})}
\newcommand\etal{{\it et al.\/}}
\newcommand\fv{{2000~FV$_{53}$}}
\newcommand\newbf{{2003~BF$_{91}$}}
\newcommand\newbg{{2003~BG$_{91}$}}
\newcommand\newbh{{2003~BH$_{91}$}}
\newcommand\logl{{\cal L}}

\begin{document}

\journalinfo{CVS \$Revision: 2.3 $ $ \$Date: 2004/04/14 $ $}
\submitted{Accepted to AJ}

\title{The Size Distribution of Trans-Neptunian Bodies\protect\footnotemark[1]}

\author{G. M. Bernstein, D. E. Trilling}
\affil{Dept. of Physics \& Astronomy, University of Pennsylvania, 
David Rittenhouse Lab, 209 South 33rd St., Philadelphia PA 19104}
\email{garyb@physics.upenn.edu}
\author{R. L. Allen}
\affil{Dept. of Physics \& Astronomy, University of British Columbia,
Vancouver BC V6T 1Z1}
\author{M. E. Brown}
\affil{California Institute of Technology, MS 150-21,
Pasadena, CA 91125}
\author{M. Holman}
\affil{Harvard-Smithsonian Center for Astrophysics, MS 51,
Cambridge MA 02138}
\and
\author{R. Malhotra}
\affil{Department of Planetary Sciences, University of Arizona, 
1629 East University Boulevard, Tucson, AZ 85721-0092}

\protect\footnotetext[1]{Based on observations made with the NASA/ESA Hubble
Space Telescope, obtained at the Space
Telescope Science Institute, which is operated by the Association of
Universities for Research in Astronomy, Inc., under NASA contract NAS
5-26555. These observations are associated with program \#GO-9433}

\begin{abstract}
We search 0.02 deg$^2$ of the invariable plane for trans-Neptunian
objects (TNOs) 25~AU or more distant using the Advanced Camera for Surveys
(ACS) aboard the Hubble Space Telescope.  With 22 ksec per pointing,
the search is $>50\%$ complete for $m_{606W}\le29.2$.  Three new
objects are discovered, the faintest with mean magnitude $m=28.3$
(diameter $\approx25$ km),
which is 3~mag fainter than any previously well-measured Solar System
body.  Each new discovery is verified with a followup 18~ksec
observation with the ACS, and the detection efficiency is verified
with implanted objects.  The three detections are a factor $\sim25$
fewer than would be expected under extrapolation of the power-law
differential sky density for brighter objects, $\Sigma(m)\equiv
dN/dmd\Omega\propto 10^{\alpha m}$, $\alpha\approx0.63$.  Analysis of
the ACS data and recent TNO surveys from the literature reveals
departures from this power law at both the bright and faint ends.
Division of the TNO sample by distance and inclination into
``classical Kuiper belt'' (CKB) and ``Excited'' samples reveals that
$\Sigma(m)$ differs for the two populations at 96\% confidence, and
both samples show departures from power-law behavior.  A 
double power law $\Sigma(m)$ adequately fits all data.  Implications of these
departures include the following.
(1) The total mass of the
``classical'' Kuiper belt is $\approx0.010M_\oplus$, only a few
times Pluto's mass, and is predominately in
the form of $\sim100$~km bodies (barring a secondary peak in the mass
distribution at $<10$~km sizes).  The mass of Excited objects is
perhaps a few times larger. (2) The Excited class has a shallower
bright-end magnitude (and presumably size) distribution; the largest
objects, including Pluto, comprise tens of percent of the total mass
whereas the largest CKBOs are only $\sim2\%$ of its mass.
(3) The derived size distributions predict that the largest
Excited body should be roughly the mass of Pluto, and the largest CKB
body should have $m_R\approx20$---hence Pluto is feasibly considered
to have originated from the same physical process as the Excited TNOs.
(4) The observed deficit of small TNOs occurs in the size regime where
present-day collisions are expected to be disruptive, suggesting
extensive depletion by collisions.  The Excited and CKB size
distributions are qualitatively similar to some numerical models of
growth and erosion, with
both accretion and erosion appearing to have proceeded
to more advanced stages in the Excited class than the CKB.
(5) The lack of detections of distant TNOs implies that,
if a mass of TNOs comparable to the CKB is
present near the invariable plane beyond~50 AU,
that distant population must be composed primarily
of bodies smaller than $\approx40$~km.
(6) There are too few small CKBOs for this population to be the
reservoir of Jupiter-family-comet precursors without a significant
upturn in the population at diameters $<20$~km.  With optimistic model
parameters and extrapolations, the Excited population could be the
source reservoir.
Implications of these discoveries for the formation and evolution of the 
outer Solar System are discussed.
\end{abstract}


\section{Motivation}
The nebular hypothesis for the formation of planetary systems is nearly
250 years old \citep{Kant} and yet observational support for the model is
relatively recent.  In the standard scenario, solids in the disk
surrounding the protostar begin to coagulate into macroscopic objects,
which accrete to kilometer sizes.  When the planetesimals become
massive enough for gravitational focusing, runaway accretion begins.
In the oligarchic growth phase, accretion is limited by excitations in
the population induced by the largest few objects.
In a protoplanetary disk, these largest planetesimals can
reach
a few $M_\earth$, sufficient to trap the
nebular gas, and rapid growth of gas giant(s) can ensue.  The nebular gas
is cleared by the stellar wind, and the remaining planetesimals are
scattered away by the giant planets.  

Today we have many observations of dust and gas disks around young stars
\citep{OB97,Beck00}, evidence that supports the nebular hypothesis.
Additionally, observations of dust disks around
somewhat older stars suggest the presence of a population of
dust-producing planetesimals in those systems, {\it e.g.} 
\citet{ST84, Gr98, Ko01}.  Some of these dust disks exhibit
structures that can perhaps be ascribed to embedded planetary systems 
\citep{KH03}. There is also now abundant evidence
for the final stage of accretion---planet formation---as extrasolar
giant planets have been detected by radial velocity and transit
observations \citep{MCM00}.  Though the basic idea of the nebular
hypothesis remains intact, each new round of observations has led to
fundamental changes in our view of planet formation.  The presence of
gas giants at $<1$~AU, for example, was not well anticipated by
theory, and migration is now recognized as an important process.

It is unfortunate that direct observation of $<1000$~km planetesimals 
in extrasolar systems is currently infeasible and likely to remain so
for many decades.  Such observation would likely reveal further
failures of imagination in our modeling of the planetesimal phase.
Fortunately, a
portion of the Sun's planetesimal population is preserved for our
examination in the region beyond Neptune, where growth time
scales are longer, the accretion process apparently did not
proceed to formation of planets, and the influence of the
giant planets was not sufficient to remove all small bodies.  Study of
trans-Neptunian objects (TNOs) provides ``ground truth'' for
models of the accretion, collisional erosion, and dynamical evolution
of planetesimal populations.  True to form, the TNO population only
vaguely resembles the preconception of a dynamically pristine 
planetesimal disk.  
With $>800$ TNOs discovered between 1992 and the present, it is clear
that the TNO population has several distinct dynamical components,
all of which appear to have eccentricity and inclination distributions
that are too broad to be the undisturbed remnants of the primordial
population.  The TNO population contains unmistakable signatures of
interactions with Neptune and perhaps other massive bodies.  With
further study we can hope to understand the dynamical history of this
region. 

The physical properties of the TNOs, particularly the size
distribution, are indicative of the accretion
process.  Observations to date are
consistent with a distribution of diameters $D$ that is a power law,
$dN/dD\propto D^q$, with $q=4.0\pm0.5$ \citep[TJL]{TJL}.  This
distribution must fail at some $D>0$ to avoid a divergence in the mass
or reflected surface brightness of the trans-Neptunian cloud, but
the scale of the breakdown in the power law is not usefully bounded by
these constraints \citep{KW01}.  
In the current dynamical environment, TNO collisions are erosive for
objects with diameters $\lesssim100$~km, so
that small objects have been removed from the population since the
events or processes that excited the TNO dynamics \citep{S95}.  Rather
soon after the discovery of the Kuiper Belt, there was speculation
that the size distribution might break at $\sim50$~km sizes
\citep{WL97}, but observations to date have not evidenced this
phenomenon.  A generic 
prediction of accretion/erosion models is a break to a shallower size
distribution below some size, but the size break is dependent upon
factors such as the duration of the accretion epoch \citep{FDS}.  
The
mass in the trans-Neptunian region must have been substantially larger
in the past in order to support the migration of Neptune 
\citep{HM99} and accretion
of the present TNO population \citep{S95}, but the relative importance
of scattering and collisional grinding in mass removal is unknown.

Extending our knowledge to the faintest (and hence smallest) possible
TNOs is clearly desirable as there may be signatures of the
collisional evolution or processes unanticipated by present theory.
It is of further interest to see if the size distribution has
dependence upon dynamical properties, as this can provide further
insight into the dependence of the accretion/collision process upon
the dynamics of the parent population.

The Hubble Space Telescope (HST) is currently the observatory of
choice for detection of the faintest possible point sources.  A
detection of a very high density of $m_V>27.8$ TNOs using the WFPC2
camera on HST is reported by \citet{CLSD}[CLSD]. For various reasons it is
likely that these detections were merely noise
\citep{BKL,Gl98} (see \citet{BAT} for further analysis of the
WFPC2 results).  The installation of the Advanced Camera for Surveys
(ACS) on HST improved the field of view, efficiency, and sampling
substantially.  This paper describes the results of a large investment
of HST time (125 orbits) into a search for TNOs using the ACS camera.

Detection of faint objects requires long integration times, but a
typical TNO moves the width of the HST point spread function (PSF) in
only a few minutes.  The ACS survey therefore uses a technique we call
``digital tracking,'' in which a long series of exposures is acquired,
with each individual exposure short enough to avoid trailing losses.  The
short exposures are shifted to follow a candidate TNO orbit and then
summed, yielding an image with long exposure time that will detect
TNOs on the chosen orbit with no trailing.  The summation must be
repeated for all plausible TNO orbits that diverge by more than the
PSF over the 
time span of the observations.  This computationally intensive
technique has been used successfully for several ground-based
faint-TNO searches \citep{TBGH, ABM01, Gl98, CB99} and a variant used by
CLSD with WFPC2 data.  We are able to detect TNOs to the fundamental
limits set by photon noise in the 22~ksec total exposure time of each
ACS search field.  The survey is $>50\%$ complete for
$m_{606W}<29.2$~mag, which is 2~mag fainter than any successful
published TNO survey and 1.5~mag deeper than the onset of false positives
in the CLSD data.  The area covered by the search is 0.02~deg$^2$,
13 times the area of the CLSD search.  The lessons learned from the
ground-based and CLSD digital-tracking surveys have helped us to
produce results that we believe are optimal and reliable.

The concepts and fundamental limits of digital tracking in this and
other applications are detailed in \citet{BAT}.  This paper
summarizes the methodology of the ACS search, presents the detections
and efficiencies, derives bounds on the apparent magnitude
distribution of the TNOs and some dynamical subsamples, and discusses
the implications for the evolution of the TNO system.
\citet{lcpaper} present the variability data for the objects
detected in the ACS survey.   \citet{astrometry} examine the current
state of the art in astrometry for moving objects, and 
the utility of high-precision astrometry for orbit determination.

\section{Detection Techniques}
The search for moving objects to the photon-noise limit of a 22,000~s
ACS integration requires a sophisticated analysis, attention to
detail, approximately 30,000 lines of code, and several CPU-years'
worth of computation on 2.4~GHz Pentium processors.
The unique tools of this data reduction are described in detail in
\citet{BAT}, but we summarize here the aspects that are important
for understanding the results.

\subsection{Observations}
The survey covers 6 slightly overlapping fields of view of the ACS
camera.  The 
spacecraft is oriented so that detector rows and columns are aligned
to the local ecliptic cardinal directions.  The 6 pointings are
arranged in a $2\times3$ mosaic, with the long axis in the ecliptic
N-S direction.  The southern two pointings are labeled A \& B, the
central two C \& D, and the northern two E \& F.  The ACS pixel scale
is nominally 0\farcs050, and nominal coverage of the full mosaic FOV
is $\approx400\arcsec\times600\arcsec=0.019\,{\rm deg}^2$.  
The exposures at a given pointing are dithered by non-integral pixel
steps, up to a few pixels, in order to improve the sampling of the
static sky objects.  The imaged field is not contiguous because our
dithers do not span the gap between the two ACS CCDs.

The field location was chosen subject to a number of criteria.  The
field center, 14h07m53\fs3 -11\arcdeg21\arcmin38\arcsec\ 
(J2000), is only 3\arcmin\ from the
invariable plane.  The field trails Neptune by 99\arcdeg, within the
libration region for perihelia of TNOs in 2:1 and 3:2 resonance with
Neptune \citep{Ma96, CJ02}.  A known TNO, \fv, is
within pointing A for the full observing period, allowing us to verify
our navigation and orbital calculations.  The field is placed and the
observations timed to minimize the loss of observing time to moonlight
and South Atlantic Anomaly crossings, and to place the field
88\arcdeg\ from opposition at the start of the observing sequence (see
below). 

All exposures were taken through the F606W filter of the ACS using the
Wide Field Camera (WFC).  In the
period UT 2003 January 26.014--31.341, which we call the ``discovery
epoch,'' $55\times400$~s exposures were taken at each of the six
pointings.\footnote{Exposure times varied slightly due to spacecraft
  constraints.} During 2003 February 05.835--09.703, the ``recovery
epoch,'' an additional $40\times400$~s exposures were taken at each
pointing.  The two sets of observations, 88--83\arcdeg\  and
77--73\arcdeg\ 
from opposition, are chosen to straddle the transition from prograde
to retrograde motion for most TNOs.  Hence any discovered objects have
maximal chance of remaining in the mosaic FOV for the full 15-day
duration of the HST observations, and the image trailing due to
apparent motion is minimized.

Individual {\em exposures} are 340--410~s long, averaging 400~s.  Five
exposures fit into a typical HST {\em orbit}, with fewer during
radiation-impacted orbits.  A set of 10 or 15 exposures is taken
during each HST {\em visit} to a given pointing.  The pointings are
visited in the pattern ABAB-CDCD-EFEF-ABAB-CDCD-EFEF during each of
the two {\em epochs} of observation.  So pointing C, for example, is
sampled sporadically, at intervals as close as 8~minutes, over a time
span of approximately 24~hours, during the first CDCD set of visits.
Approximately 2 days later, the CDCD set of visits are repeated.  Then
$\approx7$ days later, the cycle repeats for the recovery epoch.

A few shorter exposures of the six pointings and of the outskirts of
47~Tucanae are taken in order to map the WFC point spread function
(PSF) and provide astrometric tie-ins.  The performance of HST and ACS
during the observations was nearly flawless.  Comparison of the 47
Tucanae images before and after the TNO observing cycle showed
negligible change in the PSF, so we use a time-invariant (but
spatially dependent) PSF map.

\subsection{Preprocessing and ``Bright'' Object Detection}
Once the data are placed in the HST archive, they are prepared for the
moving-object search as follows:
\begin{enumerate}
\item Bias removal, flat-fielding, and bad-pixel flagging are done by
the HST ``on-the-fly'' processing.  Engineering keywords
are checked for guiding errors or other problems.  The uncertainty
images are corrected for some errors in the STScI pipeline, and we
create a weight image with the value $w$ at each pixel being
$1/\sigma^2$ (where $\sigma$ is the pixel's flux uncertainty).  The
weight is
zeroed for defective and saturated pixels, and the data and
weight images are changed into flux units.  
\item Objects in individual exposures are cataloged using {\tt
SExtractor} \citep{sextractor}.
\item The 47~Tucanae exposures are used to produce a map of the PSF
for the WFC \citep{BAT}.
\item WFC distortion maps from STScI or from \citet{Anderson} are used
to transform pixel positions into a local tangent plane for each
exposure, to accuracy $\approx10$~mas; a translation plus linear
transformation are derived for 
each exposure to register all the cataloged objects onto a global
tangent-plane coordinate system centered on the mosaic center.
\item All exposures from the discovery epoch are combined into a deep
image of the fixed sky.  This template image has 0\farcs025 pixels that
are square (no distortions) on the global tangent plane, so that the
PSF is now sampled near the Nyquist density.  Because there are 55
exposures per pointing in the discovery epoch, each template pixel has 10
or more contributing images, and the template noise level is well
below the individual exposures'.  Sigma-clipping eliminates cosmic
rays and bright moving objects from the template images.
\item We interpolate the template image to the location of
each pixel of each individual exposure.  The interpolated template is
then subtracted from each exposure.
At pixels with very high flux (centers of bright stars and galaxies),
we zero the weight image because the residuals to the template
subtraction will rise above the noise.  Note that the individual
exposures have not been resampled in producing these ``subtracted images.''
\item Artificial TNOs are added to the subtracted images.  One of
us (MH) produces a list of objects with orbital elements and
light-curve parameters selected at random from a chosen range.  The
positions, magnitudes, and motions of these objects are calculated for each
exposure.  The position-dependent PSF is trailed for the motion and
each artificial object added into the subtracted images, with
appropriate Poisson noise in each pixel.  One-third of the objects on
the list are later revealed to the searchers (GMB and DET) for use in tuning
the search algorithms.  The searchers remain blind to the other
two-thirds of the artificial objects until after a final TNO candidate
list is produced.
\item The subtracted images are searched for potential bright TNOs as
follows. 
A PSF-matched, compensated filter is scanned across each
subtracted image.  Using the weight image, we can calculate the
significance $\nu$ ({\it i.e.\/}, the signal-to-noise ratio) of each
candidate point source peak in the subtracted image.  All peaks with
$|\nu|\ge3.5$ are noted and the $\chi^2$ of a fit to the PSF is
calculated.  Those that sufficiently resemble the PSF are recorded to
a file of bright TNO candidates, to be examined later.  
Note that real TNOs will not fit the
PSF precisely because of trailing, so our criterion for matching the
PSF is kept loose, and the vast majority of candidates are cosmic
rays. \label{brightstep}
\item The subtracted images are ``cleaned'' in preparation for the
faint-object search as follows.
Every pixel in the subtracted image that deviates by more than
$5\sigma$ from the mean sky level is flagged.  All weights are set to
zero within a 2-pixel radius of each flagged pixel.  This effectively
masks all cosmic rays and non-Gaussian noise in the subtracted images,
which is extremely important for avoiding false-postive detections in
the faint-object search.  This process also masks bright TNOs and
asteroids; the former have already been detected, however, in the
previous step.\label{maskstep}
\item A ``flux image'' is now created for each exposure.  The flux
image is created on a regular grid in the global tangent-plane
coordinates.  Each such grid point is mapped back to a pixel position
on the masked subtracted image, and we record the best-fit PSF flux and
its uncertainty for a point source at that location.  Hence the ``flux
image'' is a map of the brightness of a potential point source at any
location in that exposure, and a weight (uncertainty) image is propagated
as well.  These flux images are the raw material for the faint-object
search. \label{laststep}
\end{enumerate}

Any potential bright moving objects must now be found on
lists produced in Step~\ref{brightstep}, because the masking in
Step~\ref{maskstep} may preclude their later detection.  ``Bright'' in
this context means detectable at $\ge3.5\sigma$ in a single 400~s HST
exposure, which in 
practice corresponds to $m\le 27.6$.  We use here and henceforth the
HST F606W magnitude system unless otherwise noted.  The filter
passband is roughly the union of $V$ and $R$ passbands, and the AB
zeropoint is similar to a $V$ zeropoint.

Over 900,000 flux peaks trigger the $\nu\ge3.5$ threshold in the
discovery epoch.  To fish the real (and implanted) TNOs from this sea of
cosmic rays, we first require that a flux peak repeat in the same sky
location (to 0\farcs2) on successive exposures in one orbit, leaving
7700 pairs of detections.  We reject linked detections that occur on
the same detector pixels to avoid CCD defects. 
We next require {\em two} pairs of
detections to exist within the same visit and be within 
$\approx2\arcsec\,{\rm hr}^{-1}$ of each other,  leaving 1300 candidate
quadruples of detections.  Next a preliminary orbit is fit to each
quadruple, and the methods of \S\ref{refine} used to check whether the
subtracted images are consistent with a point source moving on the
putative orbit.  This reduces the candidate list to 49 objects, of
which 46 are then revealed to be on the artificial-object list.  The
detection efficiency of the bright search for artificial objects is
found to be 100\% for $m\lesssim27.6$.  

The 3 remaining objects are
real: one is \fv, the previously known object, which at
$m=23.4$ is blindingly bright here, appearing at $\nu\approx 80$ in
each of the 55 discovery epoch exposures and 40 recovery epoch
exposures.  The second bright detection is a new object, 
now given the preliminary designation \newbg, with
time-averaged magnitude $\langle m \rangle= 26.95\pm0.02$.  The third
detection from the bright search, \newbf, has $\langle m \rangle=
28.15\pm0.04$, but is highly variable and rises above the $\nu=3.5$
single-exposure threshold several times.

The bright-object search is executed independently on the recovery
epoch observations, revealing the same three objects, which are thus
undoubtedly real.

\subsection{Faint Object Search}

The search for moving objects that are below the single-exposure
detection threshold is much more computationally intensive.  We must
sum the available exposures along any potential TNO path through the
discovery epoch exposures, then ask whether the best-fit flux for this
path is safely above the expected noise level. 

\subsubsection{The Search Space}
The space of TNO orbits is 6-dimensional, with one possible
parameterization being $\{\alpha,\beta,d,\dot\alpha, \dot\beta, 
\dot d\}$, where $\alpha$
and $\beta$ are the angular position relative to
the center of the mosaic
at some reference time $T_0$, $d$ is the geocentric distance at 
$T_0$, and the dots denote the TNO's space velocity in the same basis
({\it cf.} \citet{BK00}).
The line-of-sight motion $\dot d$ has negligible observable effect
over the course of the 15-day HST observation, so we may set it to
zero in our searches.  This means we have 5 dimensions of TNO orbit
space to search.  We search on a grid of points in this space.  The
grid spacing in $\alpha$ and $\beta$ is the pixel scale $P$ of the
flux images discussed above.  The grid spacing $\Delta v$ in the
velocity space $(\dot\alpha, \dot\beta)$ should be fine enough that
tracking errors are held to less than 1 pixel: $\Delta v \le P /
\Delta T$, where 
$\Delta T$ is the time span of the observations being combined.
Finally we must choose a grid in distance $d$.  The primary effect of
$d$ upon the apparent motion of the TNO is from the reflex of the
Earth's orbit around the Sun (and the HST's orbit around Earth).
The reflex motions scale as $1/d$, so we choose a grid that is
uniform in $\gamma\equiv 1/d$.  We also note that the non-linear
components of the TNO apparent motion all depend solely on
$\gamma$---primarily the reflex of Earth's orbital acceleration, but
also the Newtonian gravitational acceleration of the TNO itself.
The spacing $\Delta\gamma$ must be fine enough that errors in these
non-linear motion components are held to $\ll P$.  

The number of grid points that must be searched then scales roughly as
$P^{-5}\Delta T^3$.  We conduct our faint-object search in two passes:
first, with $P=0\farcs050$, and then a finer pass with $P=0\farcs030$.
The first pass runs quickly enough to be completed between receipt of
the HST data in mid-February and scheduled followup observations at
the Keck \& Magellan telescopes in late April (see \S\ref{followup}).
But the PSF of the WFC is only 0\farcs05 across, so mis-tracking by
$\approx0.5P$ at $P=0\farcs05$ causes significant blurring of the PSF
in digitally tracked images, degrading our magnitude limit by
$\approx0.2$~mag.  Hence we later run the finer-grid search to reach
the ultimate limit of the WFC data.

The bounds of the search space are determined as follows:
\begin{itemize}
\item We search $25\,{\rm AU}<d<\infty$.  Even objects at
$d\approx1000$~AU would move several ACS pixels over the course of our
  visits.
\item The perihelion of the orbit is constrained to be $\ge10$~AU.
This places a lower limit on the transverse motion at a given $d$.
\item The orbit is assumed to be bound.
This places an upper limit on the transverse motion at a given $d$.
\item The inclination of the orbit is assumed to be $i<45\arcdeg$.
This bounds the vertical component of the apparent motion.  Note we
search only prograde orbits.
\end{itemize}

\subsubsection{Steps for the Faint Search}
\label{refine}
The faint search proceeds after Step~\ref{laststep} above as follows
for each the coarse $P=0\farcs05$ (discovery and recovery epochs)
and the fine $P=0\farcs03$ (discovery only) search:
\begin{enumerate}
\item The flux images produced for this $P$ in the search are
split into six {\em sets} of visits.  Set 1 contains the
first ABAB sequence, set 2 the first CDCD sequence, etc.  The
digital-tracking sums will be accumulated over a set's worth of images, with
time span $\Delta T\lesssim 24$~hours.  Digital tracking over the full
5-day time span of the discovery epoch would be computationally infeasible.
\item For each set, the outermost loop is over the distance grid.  The
next inner loop is over the velocities $\dot\alpha$ and $\dot\beta$.
At a given distance and velocity, we calculate an orbital shift for each
exposure relative to the first exposure.  The inner loops consist of
summing the individual flux images at each pixel, with integer-pixel
shifts defined by the velocity and distance. 

In the fine search, there are 13 distance grid points and a total of
$\approx7\times 10^5$ velocity grid points in the $\{\dot\alpha,
\dot\beta, d\}$ space.  For each set there are
two pointings spanning $\approx1\times10^8$ pixels in the flux images, with
25--30 exposures per pixel per set.  In total, the fine search tests
$\approx10^{14}$ points in the TNO phase space, requiring
$\approx10^{16}$ pixel additions to do so.  This takes several
CPU-years for 2.4~GHz Pentium 4 processors, but a cluster of 10 CPUs
at Penn and 8 at Arizona reduces the required real time.

\item At each grid point of the TNO search, the point-source fluxes 
along the track of the putative TNO from all
  exposures in the set are summed, as weighted by their inverse
  uncertainties, to form a total best-fit flux and uncertainty.  If
  the significance $\nu \equiv f/\sigma_f$ exceeds a threshold of 4.0,
  the grid point is saved.

\item Above-threshold grid points that abut in phase space are
  aggregated, and the most significant is saved.  The output of the
  fine search is a list of $\approx 1.5\times10^8$ significance
  peaks in the TNO phase space.

\item For each detected peak a ``tuneup'' program is run, which fits
a model moving point source to the pixel values in
  postage stamps from all subtracted images.
  The gridded peak is the starting point and $\alpha,
  \beta, \dot\alpha$, and $\dot\beta$ are allowed to vary.  The
significance of 
  real (or implanted) objects typically rises after tuneup since the
  optimized orbit is a better fit than the nearest grid point, and the
  position/velocity estimates become more accurate.
  Significance peaks that are noise tend, however, to to become less
  significant, to have poor $\chi^2$, and/or to fail to converge.  The
  tuneup step reduces the number of $\nu\ge4.0$ peaks in the fine
  search to $6\times10^7$.

\item The tuned-up peak catalog from one set is now compared
  to all other sets of the epoch; any pair of peaks
  that might correspond to a common orbit are linked and passed to the
  next step.  Note that TNOs which cross the boundaries of the ACS
pointings are found as efficiently as those which do not.
There are $3\times10^5$ (non-unique) linked peak pairs
  in the fine search, of which $3\times10^4$ have total significance
$\nu\ge7$. 

\item All of the linked pairs with $\nu\ge7$ are again run through the tuneup
  program, but this time all of the exposures from the entire epoch
  are used.  The arc is now sufficiently long (typically 3
  days) that we can allow the distance $d$ to vary without fear of
  degeneracy.  A few detection candidates with $\chi^2-{\rm DOF}>150$
in the fit of the moving-source model to the data are
rejected; inspection shows these to be spurious detections near the
residuals of diffractions spikes of bright stars.  We apply a
threshold of $\nu\ge8.2$ ($\nu\ge10$ for the coarse search) to obtain
the KBO candidate list of $\approx 100$ objects.
\end{enumerate}
The histogram of detections vs significance rises very rapidly below
$\nu=8.2$, which is to be expected from Gaussian noise in
a search of $10^{14}$ or so phase space locations \citep{BAT}.  The
threshold is placed at the tail of this false-positive distribution.
Detection candidates above this threshold are inspected by eye, with
2--3 being clearly associated with subtraction residuals and other
data flaws.

In the coarse search, there are 92 detections with $\nu\ge10$.
The blind list of artificial TNOs is then
revealed, and 89 of 92 detections are found to be implanted
objects. Two of the remaining detections, \fv\ and \newbf,
coincide with bright detections.  The last is a new object, \newbh, 
discovered with significance $\nu=16.7$ and mean magnitude $m=28.35$.

For the fine search (which had an independent set of implanted
objects), there are 67 detections, of which 64 are found to be on the
list of implanted KBOs.  The three remaining detections are again 
\fv, \newbf, and \newbh.

The faint search technique is also applied to the recovery epoch
with a coarse ($P=0\farcs05$) grid.
The same candidates are independently detected above the
$\nu=10$ threshold.

\begin{figure*}
\plotone{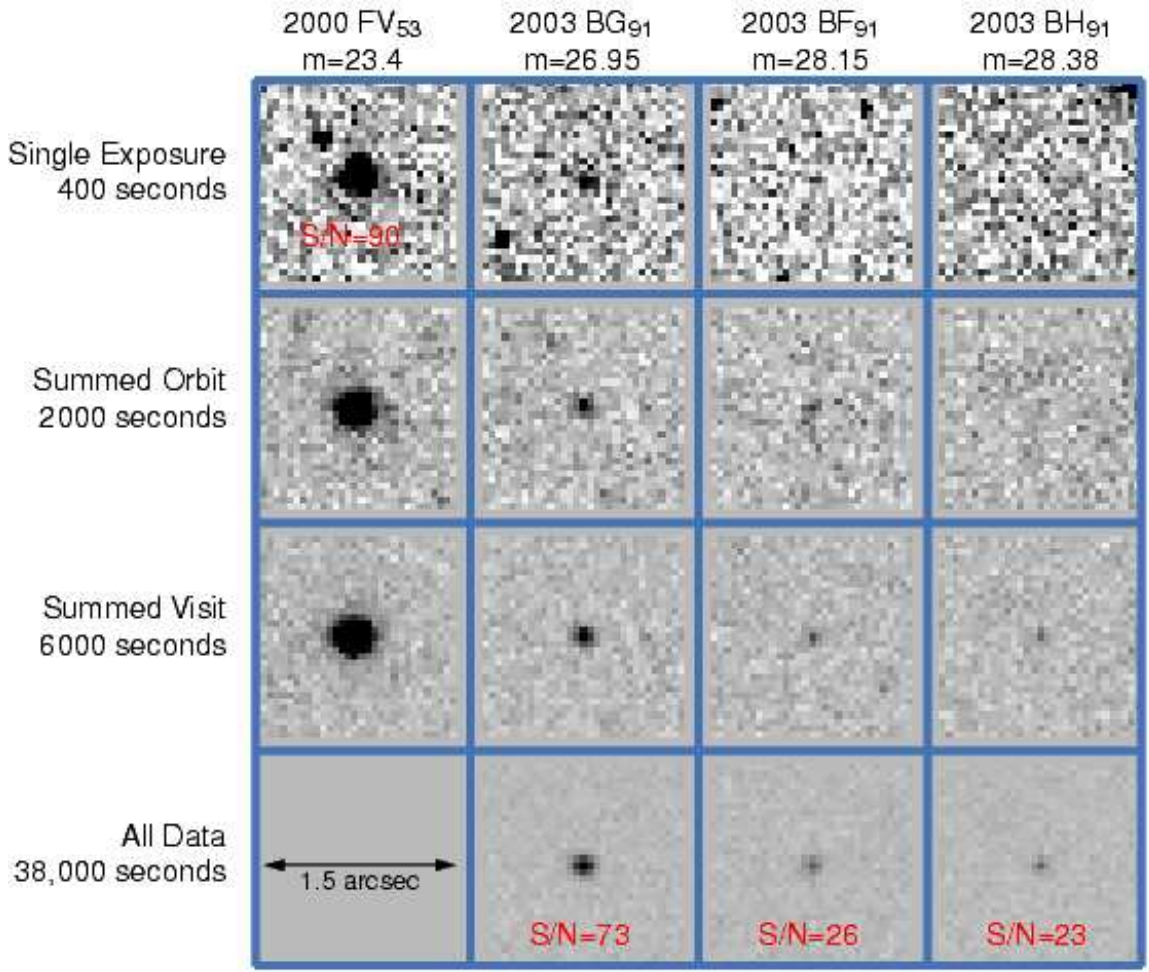}
\caption[]{\small
Postage-stamp digitally-tracked images of all four TNOs detected in
the ACS data. Successive rows show images with more contributing
integration time, starting with a randomly selected single exposure
and ending with the summed image of all available ACS data.  All
images are shown with the same greyscale. Lowest
row gives final signal-to-noise ratio of each object, save \fv, for
which the full-survey sum is omitted and the $S/N$ per exposure is
listed.  Note that the faintest object is undetectable on single
exposures, and yet is 0.8~mag brighter than our estimated completeness
limit. 
}
\label{mosaic}
\end{figure*}

Figure~\ref{mosaic} shows postage-stamp images of \fv\ and the three
new detections, as we improve the depth of images by summing more
exposures.

\subsection{Orbit Determination and Recovery}
Each of the TNOs detected in the discovery epoch is also clearly
detected in the recovery epoch.  We now combine the information from
all exposures in the entire ACS campaign to get the best possible
constraint on each object's orbit.  We again invoke the tuneup
program, whereby the orbital parameters are varied to maximize the
significance of the detection of the moving point source.  More
specifically, the orbital parameters determine the location of the PSF
and the degree of trailing in each individual exposure.  The two
endpoints of the trail are converted into pixel coordinates using the
registration information and the distortion maps.  We calculate the
PSF at the TNO location using the spatially varying PSF maps from
47~Tucanae, and we smear this PSF to the required trail length.  The
flux of the TNO is allowed to vary in a stepwise fashion from orbit to
orbit (or from exposure to exposure for the high-S/N \fv)
A model with constant flux for a given TNO would be a poor fit, as all
of the detected objects have significant flux variations.  The moving,
variable-flux model is then fit to the subtracted images, with all
orbital elements and fluxes being optimized. A byproduct
of this orbital optimization is an optimally measured light curve for
each object.  Analysis and interpretation of these light curves is in
\citet{lcpaper}.

For the final orbit determination, all six orbital parameters are
allowed to vary.  The \fv\ data are of such high quality---positional
accuracy of $\sim1$~mas for each of the 95 exposures---that the
line-of-sight velocity, and hence $a$ and $e$, are significantly
constrained with only a 13-day arc.  \citet{astrometry} will consider
in detail the techniques, limitations, and benefits of such
high-precision astrometry for the determination of Solar System
orbits. 

For the three newly detected objects, the line-of-sight motion is still
poorly determined over the 13-day arc.  In the final orbit fit to the
HST data, we include a prior constraint on the kinetic and potential
energies that weakly pushes the orbit to circularity:
\begin{equation}
\chi^2_{\rm prior} = 4(2{\rm KE}/{\rm PE}+1)^2.
\end{equation}
An unbound or plunging orbit is thus penalized as a $2\sigma$
deviation.  The result of the fitting process are best-fitting orbital
parameters (in the $\{\alpha, \beta, \ldots\}$ basis) for each object
{\em and} covariance matrices for each, 
which can be used as described in \citet{BK00} to give orbital
elements and position pre/postdictions with associated uncertainties.

\label{followup}
We attempted retrieval of all new objects using the imaging mode of
the DEIMOS instrument on the Keck~II telescope on the nights of 27
and 28 April  2003.  The error ellipses for all three objects fit within a
single DEIMOS field of view, so for each object we have 5 hours of
integration in the $R$ band on each of 2 nights.  We sum the Keck
exposures to follow the motion vectors predicted for the TNOs by the
HST data.  \newbg\ is detected at $R\approx27$ on each night, but
\newbf\ and \newbh\ remain below the detection threshold.

Attempts to retrieve the objects with the Magellan II telescope on 1--2
June 2003 were foiled by poor weather.

The orbital constraints are now refined using the Keck position.
Table~\ref{objects} gives the discovery circumstances and best-fit
orbital elements for each object.  They all have orbits consistent with
``classical Kuiper belt'' objects (CKBOs), with distances of 40--43~AU and
inclinations of $\le3\arcdeg$. The orbital eccentricities either are
(\newbg) or are consistent with (\newbf, \newbh) $e<0.08$.  It is interesting to
note that no Plutinos were discovered despite the fact that our
observations were in the longitude region where Plutinos reach
perihelion. Likewise no high-eccentricity or distant objects were
found.  The implications of their absence are discussed below.

\begin{deluxetable*}{lcccccc}
\tablewidth{0pt}
\tablecaption{Properties and Barycentric Elements of Objects Detected}
\tablehead{
\colhead{Name} & 
\colhead{$d$\tablenotemark{a}} & 
\colhead{$a$} & 
\colhead{$e$} & 
\colhead{$i$} & 
\colhead{Mean F606W mag} & 
\colhead{Diameter\tablenotemark{b}} \\
\colhead{(AU)} & 
\colhead{(AU)} & 
\colhead{} & 
\colhead{} & 
\colhead{(degrees)} & 
\colhead{} & 
\colhead{(km)} }
\startdata
\fv\tablenotemark{c} & $32.92\pm0.00$ & $39.02\pm0.02$ & 
$0.156\pm0.001$ & $17.35\pm0.00$  &
$23.41\pm0.01$ & 166 \\
\newbg & $40.26\pm0.00$ & $43.29\pm0.06$ &
$0.071\pm0.004$ & $2.46\pm0.00$ &
$26.95\pm0.02$ & 44 \\
\newbf & $42.14\pm0.01$ & $50\pm20$ &
$0.4\pm0.4$ & $1.49\pm0.01$ &
$28.15\pm0.04$ & 28 \\
\newbh & $42.55\pm0.02$ & $45\pm13$ & 
$0.2\pm0.7$ & $1.97\pm0.02$ &
$28.38\pm0.05$ & 25
\enddata
\tablenotetext{a}{Heliocentric distance at discovery.}
\tablenotetext{b}{Assuming spherical body with geometric albedo of 0.04.}
\tablenotetext{c}{Previously known TNO targeted for this
  study. Elements reported here are from ACS data alone.}
\label{objects}
\end{deluxetable*}

\subsection{Detection Efficiencies}
We next address the important issue of whether our TNO search is
complete ({\it i.e.} free of false negatives) and reliable (free of
false positives.)
\subsubsection{Reliability}
We are claiming to have examined $>10^{14}$ possible TNO sites in
phase space and have exactly zero false positives with $\nu>8.2$.  This
is not a trivial issue, as more than one publication claiming detection
of TNOs at $R>26$ has upon further examination been found 
\citep{BKL, Gl01}
to have
primarily false positive detections.  In the ACS program, however, the
detections are unambiguous: each of the 3 new objects is independently
detected in the recovery epoch as well as the discovery epoch.
Furthermore, the one sufficiently bright object is recovered at Keck.

\subsubsection{Completeness}

Is the search complete?  This issue is addressed primarily through the
implantation and blind retrieval of artificial TNOs.  We
implant two distinct sets of artificial TNOs into the discovery epoch
data:  one for the coarse search and one for the faint search.  
In each case, the
artificial TNO orbital elements are chosen at random from a
constrained range of the element space.  The range of elements is
carefully chosen so that the artificial objects overfill the ranges of
position, velocity, distance, and magnitude to which the search is 
sensitive.\footnote{The inclination range of artificial TNOs is
limited due to a software bug. We also do not place artificial TNOs 
at $d\gtrsim200$~AU.}
From the randomly selected elements, we can then calculate the
geometric search area by noting which objects fall into the
field of view for the requisite number of exposures.  From the final
object list in each search, we calculate the probability of detection
for objects that meet the geometric criteria.  The product of these
two is the {\em effective area} $\Omega_{\rm eff}$, which will be a
function of apparent magnitude, and could depend upon such quantities as
distance, rate of motion, and light-curve amplitude.

For the coarse search, $\approx150$ implanted TNOs have $27.3<m<29.4$ and
light-curve amplitudes up to $0.2$~mag.  In the bright search, 46 of
these are recovered, and 89 are recovered in the faint/coarse search.
From these we verify that there is no gap between the magnitude
ranges for which the bright search and the faint search are 100\%
effective.  The area lost to bright stars and galaxies is negligible
because the PSF of ACS images is very small, and is also stable,
so the fixed-sky subtraction is very successful.  The effective area
has no detected dependence upon TNO distance or velocity within our
TNO phase space search grid.  This is expected since all TNOs should
move several pixels from orbit to orbit, yet have average trailing
loss $<0.1$~mag.  

\begin{figure}
\plotone{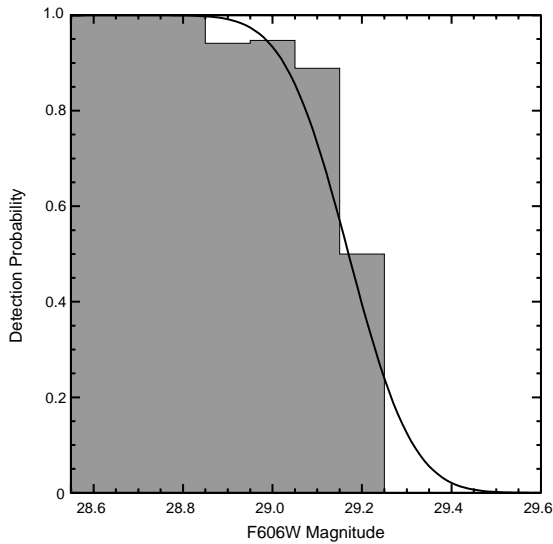}
\caption[]{\small
Probability of detection vs mean measured magnitude for the artificial
TNOs in the fine search of the discovery epoch.  The histogram gives
the results from the 101 artificial objects within the FOV, and the
curve is \eqq{aeff}.
}
\label{efficiency}
\end{figure}

For the faint search, artificial TNOs are generated with $28.6<m<29.4$
in order to more carefully probe the limiting magnitude of the
survey.  A TNO is considered to be in the survey area if it is imaged
in at least 3 of the HST visits of the discovery epoch; 101 artificial
TNOs meet this criterion, of which 64 are detected at $\nu\ge8.2$.
Figure~\ref{efficiency} plots the recovery efficiency vs mean
magnitude for the faint search.
The effective area vs magnitude is well described by 
\begin{equation}
\label{aeff}
\Omega_{\rm eff} = (\Omega_0/2) {\rm erfc}[(m-m_{50})/2w],
\end{equation}
where $\Omega_0=0.019~{\rm deg}^2$ is the peak effective solid angle,
$m_{50}=29.17$
is the F606W mag at which the effective area drops 50\%, and
$w=0.08$~mag is a transition width.  The detection efficiency again
has no measurable dependence upon distance or velocity over the search
space.

Why should we trust the artificial-KBO tests to verify our
completeness? After all, if the
implantation process and the search/extraction process make common
errors in flux scale, orbit calculation, PSF shape, or image
distortion, then the artificial objects could be detected at high
efficiency while real objects are not.

We note first that the orbit calculation code used for object
implantation was written by one of us (MH) while that used for
extraction was independently written by another (GMB), and both codes
were checked against each other and the JPL online Horizons
service.\footnote{{\tt http://ssd.jpl.nasa.gov/horizons.html}}

The targeted TNO \fv\ helps us to address concerns about errors in
orbit calculations, image registration and distortion, or PSF
estimation.  The individual exposures for \fv\ are fit by our
moving-PSF model to good precision:  its positions match the
extrapolation of previous observations to the accuracy of the
extrapolation, and we find the positions consistent with a refined
orbit to the level of $\approx 3$~mas, or $<0.1$~pixel on the WFC.
This is better than the claimed accuracy of the distortion map.  We are
thus reassured that our models and code for spacecraft navigation, image
registration, orbital motion, and field distortion are correct to the
accuracy required for the search.  The images of \fv\ are formally
inconsistent with the PSF model ($\chi^2$ per DOF is $>1$) but this is
because the $S/N$ ratio of the \fv\ observations are very high.  The
deviations from the fit are at the level of a few percent of the PSF,
and hence the PSF model is sufficiently accurate for the fainter
detections. 

\subsubsection{A Caveat on Light Curves}
\label{caveat}
The selection function for TNOs with variable magnitude is complex:
the object must be seen during at least 3, preferably 4, HST visits
with a $S/N$ ratio of $\gtrsim 4$, to survive the detection cuts.
The timing of these visits is irregular, hence there is no
simple way to quantify the impact of light-curve variation on
detectability.  The implanted TNOs were given sinusoidal light curves with
peak-to-peak amplitudes chosen uniformly between 0 and 0.2~mag and
periods chosen uniformly between 0.05 and 1.3 days---in this range, we
did not note any change in detection probability vs magnitude.  We
know, however, that there exist TNOs with light curve amplitudes near
or above 1~mag, such as \newbf\ \citep{lcpaper} and 2001~QG$_{298}$
\citep{SJ04},
and we should 
investigate the effects
of high variability on the detection properties of our and other
surveys. 
Subtle biases on light-curve shape are present for all TNO surveys,
though other authors have chosen, like us, to ignore them for
simplicity.  In Appendix~\ref{appa} we demonstrate that these biases
are too small to be significant with current data, but may be
important for future larger surveys.

\section{Constraints on the Trans-Neptunian Population}
\label{constraints}
The ACS survey detects a total of three objects (not counting the
targeted \fv), described in Table~\ref{objects}, over an effective
search area described by \eqq{aeff}.  TJL fit a power law to
the cumulative ecliptic surface-density distribution of TNOs:
\begin{equation}
\label{cumulativeplaw}
N(<R) = 10^{\alpha(R-R_0)}\,{\rm deg}^{-2},
\end{equation}
with $R$ being the $R$-band apparent magnitude,
$\alpha=0.63\pm0.06$ and $R_0=23.0$.
Their fit is to survey data from $19<R<27$.  Taking
our limit $m_{50}=29.17$ to be equivalent to a limit of
$R\le28.8$ (\S\ref{colors}), an extrapolation predicts $\approx85$
detectable objects in our survey.  This is quite
inconsistent with our observation---even the TJL $2\sigma$ limit of
$\alpha=0.51$ predicts $\approx16$ detections in our survey---and it
is immediately obvious that 
the magnitude (and hence size) distribution of TNOs changes behavior
somewhere in the $25<m<29$ range.  In this section we quantify the
nature of this breakdown in the single power law, and calculate the
implications for integral properties of the TNO population.

\subsection{Compendium of TNO Survey Data}
\label{compendium}
We wish to derive the differential surface density $\Sigma(R)\equiv
dN/dR\,d\Omega$ of TNOs per $R$ magnitude interval using all possible
reliable published survey data.
The requirements for published survey data to be useful are:
\begin{itemize}
\item The coordinates of all fields searched must be given.
\item The effective search area as a function of $m$ for each field
  must be given, 
  preferably derived by Monte-Carlo tests.
\item The circumstances of discovery of all detected objects must be
  given, including apparent magnitude, estimated heliocentric
  distance $d$, and estimated inclination $i$ of the orbit.
\end{itemize}
Note that we do not require that all detected objects have
fully determined orbits.    While
the $R\le24$ TNO discoveries have been recovered with admirable
completeness, the practical difficulties of recovery for fainter
objects have precluded observational arcs longer than $\approx 1$~day
for any object with $R>25.6$ (prior
to this ACS survey).  Fits to one-day arcs yield $d$ and $i$ to
10--20\% accuracy but other orbital properties are highly degenerate.
Hence a comprehensive study of both and bright and faint detections
can as yet make use only of $d$ and $i$ to categorize the TNOs.

The sky-plane density of TNOs is certainly a function of latitude
relative to the midplane of the population.  Most of the $R>22$
searches have been targeted to the ecliptic plane, or less frequently
the invariable plane.  Brighter surveys cover larger area and have
ranged farther from the ecliptic.  Proper comparison of bright to
faint TNO densities requires that we consider sky densities measured
within a fixed band of TNO latitude.  If the latitude distribution of
TNOs were well known we could make use of all the available survey
data.  While the midplane of the TNO population has recently been
estimated to be $\approx0.7\pm0.4\arcdeg$ from the invariable plane
\citep{BP04}, the full distribution remains poorly constrained.
We restrict the published survey
data compendium to fields with invariable latitude $\le3\arcdeg$.  The TJL
estimate of the inclination distribution of bright CKBOs implies
a drop by factor $\approx 2$ in the sky plane density
from zero to three degrees {\it ecliptic} latitude, so there may
remain substantial inhomogeneity in comparing surveys over a
$\pm3\arcdeg$ swath.  Attempts to select a
narrower latitude range are counterproductive, however, given our poor
knowledge of the sky distribution of TNOs.

The resonant TNO population has longitudinal structure in the
sky-plane density as well, with more objects being found at bright
magnitudes in the directions perpendicular to Neptune.  Plutinos (3:2
Neptune resonators) have perihelion positions that librate about these
points.  This longitude variation has yet to be mapped in any way, so
a correction is not possible.  The effect upon our results is not
likely to be significant, because surveys at all magnitudes span a
range of longitudes.  The exception is our uniquely deep ACS field,
which though pointed in the region of Plutino perihelion libration does
not detect any Plutinos.  Our most precise analyses will in any
case be done on samples intended to exclude Plutinos.

A subtle difference between the bright and faint TNO samples is that
the former are typically discovered on single short
($\lesssim10$~minutes) exposures, and hence measure the {\em
  instantanenous} magnitude distribution.  Objects with $R>25$ are
detected in summations of many hours' worth of exposures, and depend
upon the flux of the TNOs {\em averaged over their light curves}
(see also discussion in \S\ref{caveat}).  In Appendix~\ref{appa} we
show that this effect is insignificant for the current data.

\label{colors}
We list in Table~\ref{surveys} the published TNO surveys that meet the
requirements.  We restrict our consideration to those works that
dominate the surveyed area at a given magnitude, and we omit surveys
that have been shown to contain significant false-positive
contamination.  For the purposes of the $\Sigma(R)$ analysis, we wish to
standardize all magnitudes to the $R$ band.  The La and CB data are
reported in $V$ band; \citet{TR03} present accurate colors for many
(bright-end) TNOs, and the mean $V-R$ is $\approx0.6$~mag, which we
apply to the $V$ detections.  Some of the ABM fields use a ``$V\!R$''
filter, so for these we apply the color correction given by ABM,
assuming again $V-R=0.6$ for an average TNO.  The F606W filter on the
ACS WFC is essentially the union of the $V$ and $R$ passbands.
\citet{TR03} show that the average TNO is $\approx0.39$~mag redder in
$V-R$ than the Sun, so we presume that the average $m_{606}-R$ TNO color
is about 0.20~mag redder than Solar.  Taking the Solar
$V-m_{606}=0.06$~mag from the ACS Instrument Handbook, an average TNO
should have $m_{606}-R\approx0.4$~mag.  We correct the ACS limits and
detections to $R$ using this value.  Henceforth we will use only
$R$-band magnitudes.  In Appendix~\ref{appa} we show that variance in
$V-R$ colors of TNOs has negligible effect upon our analysis of the
current data.

Some other adjustments to the published survey data are necessary:
\begin{itemize}
\item The effective
  search area of each \citet{La01} field is taken to be the product of its
  geometric area and the $F(T)$ entry denoting the fraction of the
  field that is estimated to be unique to the survey in their
  Table~1.  
  We crudely fit a completeness model of the form
  (\ref{aeff}) to the completeness for each seeing bin listed in their
  Table~3.   The effective area of all search fields centered within
$\pm3\arcdeg$ invariable latitude are summed to give a total useful
survey area for the survey, and we only count objects detected in
these low-latitude fields.  The redundancy and broad latitude coverage
of this survey mean that its peak effective area, for our purposes, is
only 20\% of its raw angular coverage.
\item  In the TJL data, no distance or inclination information
  is available for 7 of 74 
  objects detected near the ecliptic plane.  TJL note that this
  information is missing because of inclement weather at followup time
  and therefore these 7 objects should be drawn from the same
  distribution as the remainder.  We therefore omit these 7 from our
  listing and decrease the tabulated effective areas by $7/74=9.5\%$ in
  order to reflect this followup inefficiency.
\item \citet{ABM01} and \citet{ABM02} [ABM] are merged for this
  analysis.
\item \citet{Gl01} describes two searches, one with CFHT and one
  with the VLT.  We sum their effective areas and detections in this
  analysis. 
\item \citet{TB} do not give individual field coordinates, but do give
the total sky coverage as a function of invariable latitude, which
suffices for our purpose.  This preliminary report does not include a
detection efficiency analysis, merely an estimate of a 50\%
completeness level $R\approx20.7$.  We avoid this uncertainty by
making use of the TB data only for $R<20.2$, and assuming that in
this range the detection efficiency is a constant 85\% over the
surveyed area.  Note that the effective search area of TB comprises
the majority of the $\pm3\arcdeg$ latitude region.
\item The brightest surveys \citep{TB, La01} have inaccurate
  magnitudes for their detections, due to varying observing conditions
  and ill-defined passbands. Nearly all of these objects have,
  however, been carefully reobserved by other authors for color and
  variability 
  information, and we can replace the original survey magnitudes with
  highly accurate $R$-band mean magnitudes.  The original magnitude
  uncertainties remain relevant, however, for treatment of
  incompleteness, as discussed in Appendix~\ref{appa}.
\item We truncate all the efficiency functions $\eta$ to zero when
they drop below $\approx15\%$ of the peak value for that survey, and
ignore detections faintward of this point.  In this way we avoid making
our likelihoods sensitive to rare detections in the (poorly
determined) tails of the detection function.
\end{itemize}

\begin{deluxetable*}{clccccccc}
\tablecolumns{9}
\tablewidth{0pt}
\tablecaption{Summary of TNO Surveys}
\tablehead{
\colhead{Abbreviation} & 
\colhead{Reference} & 
\colhead{$\Omega_{\rm eff}$ (deg$^2$)\tablenotemark{a}} & 
\colhead{$m_{50}$\tablenotemark{a}} &
\colhead{$N$ (CKBO)\tablenotemark{b}} &
\colhead{$N$ (Excited)\tablenotemark{b}} &
\colhead{$P(\le N)$\tablenotemark{c}} &
\colhead{$Q_{AD}$\tablenotemark{c}} &
\colhead{$P(\le\logl))$\tablenotemark{c}}
}
\startdata
ACS	& This work	& 0.019	& 28.7 &  3 &  0 & 0.16 & 0.65 & 0.42 \\
CB	& \citet{CB99}	& 0.009	& 26.8 &  1 &  1 & 0.98 & 0.91 & 0.09 \\
Gl	& \citet{Gl01}	& 0.322	& 25.9 &  8 &  9 & {\bf 0.98} & {\bf 0.03} & 0.83 \\
ABM	& \citet{ABM02}	& 2.30	& 25.1 & 17 & 15 & 0.49 & 0.18 & 0.58 \\
TJL	& \citet{TJL}	& 28.3	& 23.8 & 39 & 28 & 0.27 & 0.44 & 0.27 \\
La	& \citet{La01}	& 296.	& 20.8 &  1 &  5 & {\bf 0.97} & {\bf 0.05} & 0.30 \\
TB	& \citet{TB}	& 1430.	& 20.2\tablenotemark{d} &
					  0 &  2 & 0.28 & 0.63 & 0.58 \\
\enddata
\tablenotetext{a}{Effective search area within 3\arcdeg\ of the invariable
  plane at bright magnitudes, and $R$
  magnitude at which effective area drops by 50\%.}
\tablenotetext{b}{Number of detected TNOs in the two dynamical classes
  defined in the text.}
\tablenotetext{c}{Cumulative probabilities of this survey under the
  best-fit two-power-law model, for Poisson test, Anderson-Darling
  test, and $\logl$ tests, as described in text. Boldface marks
indications of poor fits.}
\tablenotetext{d}{The TB data are not used faintward of 20.2~mag.}
\label{surveys}
\end{deluxetable*}

We define three dynamical groupings of the detected TNOs in these
surveys:
\begin{enumerate}
\item The {\bf TNO} sample holds all objects discovered at
  heliocentric distances $d>25$~AU.  One known Centaur
  (1995~SN$_{55}$) sneaks into this TNO sample, but we do not omit it
  because similar objects found in the faint sample would not have
  been rejected.
\item The {\bf CKBO} sample is the subset of the TNO sample having
  $38<d<55$~AU and $i\le5\arcdeg$.  This is intended to exclude
  resonant and scattered objects to the extent possible with our
  limited orbital information.
\item The {\bf Excited} sample is the complement of the CKBO sample
  in the TNO sample.  High inclinations and/or proximity to
  Neptune would indicate substantial past interactions with Neptune or
  another massive body.
\end{enumerate}

Note that we have used ecliptic inclinations rather than
invariable, since the latter are not generally available.  We have
used the central values for $d$ and $i$ even when the surveys report
uncertainty ranges that cross our definitional boundaries.  We have
ignored the possibilities of overlaps in survey areas, and omitted
targeted objects such as \fv.

The TNO sample under analysis thus contains 129 detections spanning
$19.5\le R \le 28.0$, of which 69 are assigned to the CKBO class.
Figure~\ref{binned} shows the $\Omega_{\rm eff}$ of the published surveys
vs magnitude, and the binned magnitude distribution of the detections.

Our definition of the CKBO class is imperfect because we are
restricted to use of $d$ and $i$ in classification.  
Resonant and
``scattered-disk'' TNOs can also slip into the CKBO category under
some conditions, and Centaurs near aphelion may be accepted as either
CKBOs or Excited TNOs.  
Of the objects classed as CKBOs in this study, 39 have sufficiently
long arcs to determine $a$ and $e$.  Of these, all have $42<a<48$ and
$e<0.2$ except the $R=20.9$ Centaur 1995~SN$_{55}$ and the scattered
$R=23$ object
1999~RU$_{214}$. 

Of the 33 objects with well-known $a$ in our Excited class, 
all have $a>33$, though a few are Neptune-crossing and might be
labelled Centaurs by some authors' criteria.

Therefore if these 72 objects with good orbits are a guide, a few
percent of all objects would be classified differently if full orbital
elements were used instead of just $d$ and $i$.  

\subsection{Statistical Methods}
\label{stats}
We wish to ask what forms for the differential surface density 
$\Sigma(R)\equiv dN/dR\,d\Omega$ are most consistent with the
collected survey data.  Note that throughout this paper we will
consider the {\em differential} distribution with magnitude instead of
the {\em cumulative} distribution that is fit in most previous works.
The expected number
of detections from a perfect survey over solid angle $\Omega$
in a small magnitude interval $\Delta R$ is
\begin{equation}
\Delta N = \Sigma(R) \Omega\, \Delta R.
\end{equation}
Appendix~\ref{appa} is a detailed explanation of the form of the
likelihood $L$ of observing TNOs at a set of magnitudes $\{m_i\}$
given an assumed $\Sigma(R)$.  This is general complex if the details
of light curves, photometric errors, color corrections, 
and detection probability
must be considered.  The Appendix demonstrates
that it is safe to take a simplified approach that ignores many of
these details, which we present here.

The true surface density $\Sigma(R)$ must be convolved with: the color
conversion to the observed-band magnitude $m$; the measurment error on
$m$ due to noise and variability; the detection efficiency; and any
inhomogeneities of the survey,
leaving us with a function $g(m)$ that describes the expected
distribution of measured magnitudes in this survey.
The expected number of detections from some particular survey is
\begin{equation}
\bar N = \int\!dm\, g(m) = \int\! dR\, \Omega\,\eta(R)\Sigma(R),
\end{equation}
where $\eta(R)$ is the detection probability for a TNO of mean
magnitude $R$ that lies within the geometric area $\Omega$ of the survey.
This quantity can be determined from Monte Carlo tests.

The likelihood of observing a set of $N$
magnitudes $\{m_i\}$ under an assumed distribution $g(m)$ is
\begin{equation}
L(\{m_i\} | g)  \propto e^{-\bar N} \, \prod_{i=1}^N g(m_i).
\label{gladman}
\end{equation}
In this work we will make the approximation that the difference
between the observed magnitude $m_i$ and the true $R$ magnitude is
minimal (aside from a constant color term) so that we may
approximate
\begin{eqnarray}
\label{gapprox}
g(m) & \rightarrow & \Omega \eta(m) \Sigma(m) \equiv \Omega_{\rm eff}(m)
\Sigma(m) \\
\label{lapprox}
\Rightarrow \qquad
L(\{m_i\} | \Sigma)  & \propto & 
	e^{-\bar N} \, \prod_{i=1}^N \Omega_{\rm eff}(m_i) \Sigma(m_i) \\
\bar N & = & \int dm\, \Omega_{\rm eff}(m) \Sigma(m).
\end{eqnarray}

When fitting alternative forms of $\Sigma(R)$ to the survey data, the
one that maximizes the likelihood (\ref{lapprox}) (times any prior
probability on
the models) is the Bayesian preferred model.  Confidence intervals on
the parameters of the underlying $\Sigma(R)$ can be derived from this
probability function as well.  We will always take the prior
distributions to be uniform, with the exception that the overall
normalization of $\Sigma$ has a logarithmic prior.

\subsubsection{Goodness of Fit}
We will be producing models for $\Sigma(m)$, and hence $g(m)$, which
best fit the data.  We then ask whether the observations are in fact
consistent with having been produced by this model.  The general
approach is to define some statistic $S$ and ask whether the measured $S$ is
consistent with the range of $S$ produced by realizations of the
model.  We will test goodness-of-fit with two statistics.  The first
is simply the likelihood $L(\{m_i\}|\Sigma)$ itself, given in
\eqq{gladman}.  The probability $P(<L)$ of a
realization of the model having lower likelihood than the measurements
will be calculated by drawing random realizations from the best-fit
distribution.  Values $P<0.05$ or $P>0.95$ are signs of poor fit.

We also use the Anderson-Darling (AD) statistic, defined as \citep{NR}
\begin{equation}
AD = \int { [S(m) - P(m)]^2 \over P(m) [1-P(m)]} dP(m).
\label{ad}
\end{equation}
Here $P(m)$ is the cumulative probability of a detection having
magnitude $\le m$, so $0\le P \le 1$.  The cumulative distribution
function of the observed objects is $S(m)$.  The AD statistic is
related to the more familiar Kolmogorov-Smirnov (KS) statistic, which
is the maximum of $|S(m)-P(m)|$, but is more sensitive to the tails of
the distributions.  We calculate $Q_{AD}$, the probability of a random
realization having {\em higher} $AD$ value than the real data.  Values
of $Q_{AD}<0.05$ indicate poor compliance with the model distribution.

Because the likelihood and AD values of the real data are calculated
from a $g(m)$ that is the best fit to the data, it is necessary to
also fit each random realization before calculating $L$ or $AD$.
Because the normalization of $g$ is always a free parameter, 
we fix each random realization
of $g(m)$ to have the same number of detections as the real data.

We note further that we always sum the effective areas and detections
of all surveys before analyzing the data, rather than considering the
likelihood of each component survey.  We believe this makes the fit a
little more robust to small errors in individual surveys' detection
efficiency estimates.  In \S\ref{consistency} we examine whether each
constituent survey is consistent with the $\Sigma(R)$ derived from the
full dataset.

\subsection{Single Power Law Fits}
Previous fits to the magnitude distribution of TNOs have assumed that
the cumulative, and hence differential, distributions fit a single
power law.  We attempt to fit the ACS and previous survey results to a
differential distribution of the form
\begin{equation}
\label{plaw1}
\Sigma(R) = \Sigma_{23} 10^{\alpha(R-23)}\,{\rm deg}^{-2}.
\end{equation}
We first fit this law to the older surveys, excluding the ACS and TB
data.  We recover a best fit of $\alpha=0.61\pm0.04$ for the TNO
sample, which is consistent with the previous fits, {\it e.g.},
TJL.  This best-fit power law is a marginally acceptable fit to the data,
with likelihood probability $P(<L)=0.92$ and AD
probability $Q_{AD}=0.06$.

When we include the ACS and TB data in the power-law fit we find
the best-fit slope drops to $\alpha=0.58\pm0.02$.  The fit is strongly
excluded, however, $P(<L)=0.997$ and $Q_{AD}\le0.001$.
The probability of detecting so few objects in the ACS survey under
this power law is $P(<N)<10^{-14},$ and the TB survey is also highly
deficient.  A single power law extending to the ACS data is ruled
out at very high significance, as expected.  By contrast, there is
16\% probability of finding $\le3$ TNOs in the ACS survey
under the
best-fit double-power-law $\Sigma(m)$ (\S\ref{double}).

\subsection{Binned Representation}
\begin{figure}
\epsscale{1.2}
\plotone{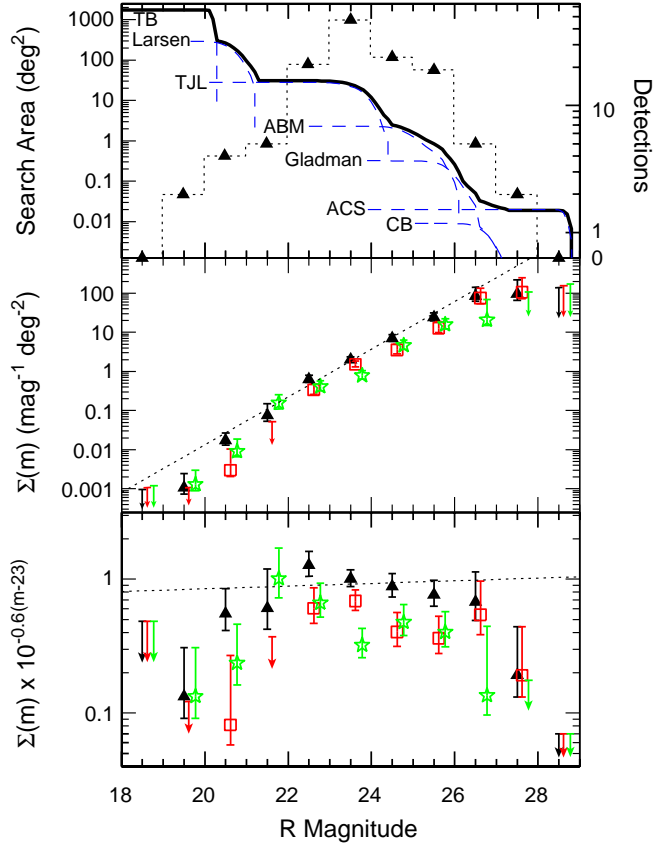}
\caption[]{\small
The top panel shows the total effective
survey area (left axis) within $\pm3\arcdeg$ invariable latitude vs magnitude,
both summed (solid curve) and for individual
surveys (dashed curves).  The histogram shows the number of detected
TNOs for the combined surveys (right axis).
The middle panel plots the binned estimate (Bayesian expectation and
68\% credible range) of the
differential TNO surface density near the invariable plane.  Solid
triangles are for the full TNO sample, open squares (red) are for the
CKBO sample, and open stars (green) are the Excited sample. The latter
two are slightly displaced horizontally for clarity.  The dashed line
is the best single-power-law fit to the older data.  The lower panel
shows the binned surface density relative to an $\alpha=0.6$ power
law (same symbols), {\it i.e.}, the $\Sigma_{23}$ values from a
stepwise fit to \eqq{plaw1}.  The departure of all samples from a simple power
law is clear.  This plot is also useful in that the vertical axis is
the mass per magnitude interval, if the albedo, material density, and distance
are independent of magnitude.
}
\label{binned}
\end{figure}

Figure~\ref{binned} presents a non-parametric, binned estimate of the
differential surface density.  The survey data for each one-magnitude interval
from $18<R<29$ are fit to the form (\ref{plaw1}), with $\alpha$ fixed
to 0.6 and $\Sigma_{23}$ free.  The expectation $\bar\Sigma_{23}$ and
the 68\% Bayesian credible regions are calculated as described in
\S\ref{marginsection}.  The middle panel shows the
resulting expectation of $\Sigma$ at the center of each bin, and the
lower panel plots the $\bar\Sigma_{23}$ values, {\it i.e.}, the deviation
of each bin from a pure $\alpha=0.6$ power law.
This plot is useful
for visualizing the departures from power-law behavior, but we always
fit models to the full survey data rather than the binned version.
It is immediately apparent that the TNO surface density departs from a
single power law at both the bright and faint ends of the observed
range, for both the CKBO and Excited subsamples.

\subsection{Rolling Power Law Fits}
As a next level of complication, we consider a surface density with a
rolling power law index:
\begin{equation}
\Sigma(R) = \Sigma_{23} 10^{[\alpha(R-23)+\alpha^\prime(R-23)^2]}.
\label{rolling}
\end{equation}
Note that this is a log-normal distribution in the flux, roughly so
for diameter as well.

\begin{figure}
\plotone{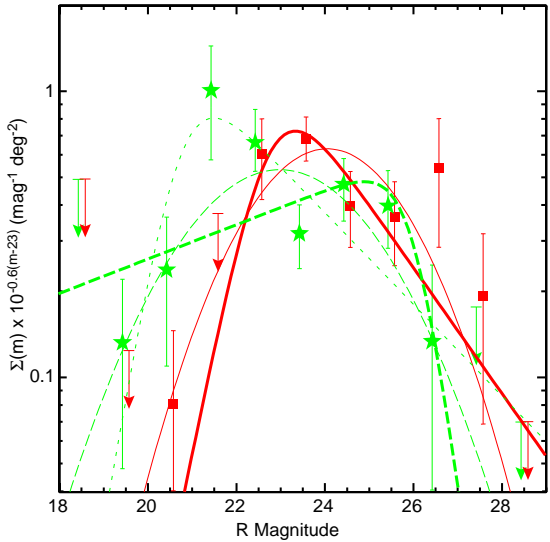}
\caption[]{\small
Best-fit models for the differential surface density $\Sigma(R)$ of
TNOs are plotted along with the binned representation of the data from
Figure~\ref{binned}. The (green) stars are binned data for the Excited
sample and (red) squares are the CKBO sample, and we have
again divided out the function $10^{0.6(R-23)}$.  The two heavy curves
are double-power-law fits, and the two thin curves are rolling-index
power laws.  Red (solid) curves are for the CKBO sample and green
(dashed) for the Excited sample. The dash-dot curve is a secondary
double-power law fit to the Excited which is consistent with these
data, but inconsistent with the existence of Quaoar-sized objects (or
Pluto).  The precipitous drop in the best-fit double power laws at the
bright (CKBO) and faint (Excited) ends is an artifact of the absence
of detections in this sample.  There are less precipitous drop-offs
that are quite consistent with the data, as is apparent from Figure~\ref{a1a2contour}.
}
\label{fitfig}
\end{figure}

This fit to the full TNO sample is now acceptable, with
$Q_{AD}=0.55$ and $P(\le L)=0.18$.  The CKBO and
Excited samples are only marginally well fit, with  $Q_{AD}=0.04$ and 0.06,
respectively. 
Best-fit parameters are given in Table~\ref{bestfits}, and the best
fit $\Sigma(R)$ are plotted over the data in Figure~\ref{fitfig}.

The addition of the single parameter $\alpha^\prime$ to the
single-power-law fit leads to highly significant improvements in the
likelihood:  $\log L$ is increased by 32, 22.2, and 12.6 for the TNO,
CKBO, and Excited samples, respectively.  This is equivalent to
$\Delta \chi^2 = 2\Delta(\log L) \ge 25$ for one additional parameter,
which has negligible probability of occurring by chance.  Hence the
single-power-law fits are strongly excluded.

\begin{figure}
\plotone{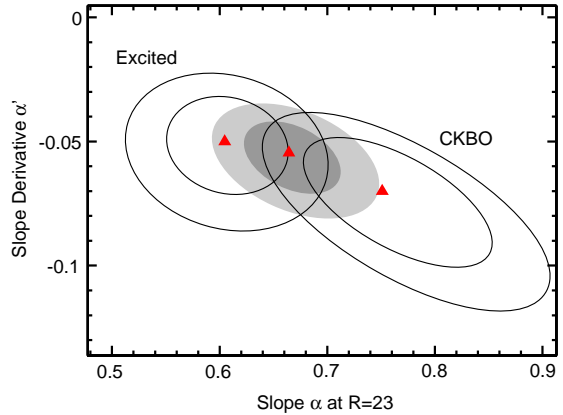}
\caption[]{\small
Allowed ranges of the slope and derivative for rolling-power-law fits
to the sky density of TNOs, as per \eqq{rolling}.  Shaded regions are
for the full TNO sample, 
while solid contours are for the CKBO and Excited subsamples.  In all
cases contours enclose 68\% and 95\% of the total posterior density.  The
curvature $\alpha^\prime$ of $\Sigma(R)$ is clearly non-zero, and
the two dynamical subsamples are distinct.  The lower $\alpha$ for the
Excited sample implies that its mean
magnitude and mass are larger than those of the CKBOs.
}
\label{rollingcontours}
\end{figure}

Using the Bayesian
approach of \S\ref{stats} we may produce a probability function
$P(\Sigma_{23}, \alpha, \alpha^\prime | \{m_i\})$ given
the observations.  
Figure~\ref{rollingcontours} plots the credible regions for
$\alpha$ and $\alpha^\prime$ in the three samples.
Note first that $\alpha^\prime=0$ is
strongly excluded, {\it i.e.} a rolling index is required.  For the
rolling-index model, any $\alpha^\prime<0$ gives convergent integrals
for TNO number and mass at both bright and faint ends.  Second we
see that the CKBO sample requires a larger $\alpha$, meaning that
$\Sigma(R)$ is steeper at $R=23$ than for the Excited sample, {\it
i.e.} the CKBO sample is shifted to fainter magnitudes relative to the
Excited sample, by about 1~mag.  In the next section we will discuss
the implications of this magnitude shift.

\begin{deluxetable*}{lrrcccrrrccc}
\tablecolumns{10}
\tablewidth{0pt}
\tablecaption{Best Fit Differential Surface Density Models}
\tablehead{
\colhead{Sample} & 
\multicolumn{5}{c}{Rolling Power Law} & 
\multicolumn{6}{c}{Double Power Law} \\
\colhead{} & 
\colhead{$\alpha$} & 
\colhead{$\alpha^\prime$} & 
\colhead{$\Sigma_{23}$} & 
\colhead{$P(\le L)$} & 
\colhead{$Q_{AD}$} & 
\colhead{$\alpha_1$} & 
\colhead{$\alpha_2$} & 
\colhead{$R_{\rm eq}$} & 
\colhead{$\Sigma_{23}$} & 
\colhead{$P(\le L)$} & 
\colhead{$Q_{AD}$} \\ 
\colhead{} & 
\colhead{} & 
\colhead{} & 
\colhead{(mag$^{-1}$ deg$^{-2}$)} & 
\colhead{} & 
\colhead{} & 
\colhead{} & 
\colhead{} & 
\colhead{} & 
\colhead{(mag$^{-1}$ deg$^{-2}$)} & 
\colhead{} &
\colhead{}  
}
\startdata
TNO	& 0.66 & -0.05 & 1.07 & 0.18 & 0.55 & 0.88 & 0.32 & 23.6 & 1.08 &
0.16 &
0.12 \\
CKBO	& 0.75 & -0.07 & 0.53 & 0.54 & 0.04 & 1.36 & 0.38 & 22.8 & 0.68 &
0.71 & 0.23 \\
Excited	& 0.60 & -0.05 & 0.52 & 0.04 & 0.06 & 0.66 & -0.50 & 26.0 &
0.39 & 0.24 & 0.13
\enddata
\label{bestfits}
\end{deluxetable*}

\subsection{Double Power Law Fits}
\label{double}
We next consider a surface density that is the harmonic mean of two
power laws:
\begin{eqnarray}
\label{plaw2}
\Sigma(R) & = & (1+c) \Sigma_{23} \left[
10^{-\alpha_1(R-23)} + c 10^{-\alpha_2(R-23)}\right]^{-1}, \\
c & \equiv & 10^{(\alpha_2-\alpha_1)(R_{\rm eq}-23)}. 
\end{eqnarray}
Under the convention $\alpha_2<\alpha_1$, the asymptotic behavior of
this function is a power law of indices $\alpha_1$ at the bright end
and $\alpha_2$ at the faint end, with the two power laws contributing
equally at $R_{\rm eq}$.  The free parameters for this
model are $\{\alpha_1, \alpha_2, R_{\rm eq}, \Sigma_{23}\}$.  
We introduce the double-power-law model for two reasons:
first, in the next section we will be interested in how strongly the
parameterization of $\Sigma(R)$ affects our conclusions, so we want
some alternative to the rolling-index model.  Second, some models for
accretion/erosion of planetesimals predict asymptotic power-law
behavior, which is absent in the rolling-index model.

The double-power-law model adequately describes the TNO, CKBO, and
Excited samples, with $Q_{AD}\ge0.12$ and $P(\le L)\ge0.16$.  The
best-fit parameters are listed in
Table~\ref{bestfits} and plotted with the binned representation of the
data in Figure~\ref{fitfig}.  

The values of $\log L$ for the double-power-law fits are within
$\pm1.3$ of those for the rolling-power-law fits.  So while the
double power law is clearly superior to the single power law, the
likelihood itself offers no preference over the rolling power law.
The Anderson-Darling statistic is, however, more acceptable for the
double than for the rolling power law fits to the CKBO and Excited
samples (Table~\ref{bestfits}).  There is weak statistical preference
and theoretical prejudice for the double power laws; in \S\ref{brightest} we
note that the rolling-power-law fits do not properly describe the
number of very bright Excited TNOs found away from the invariable plane.

\begin{figure}
\plotone{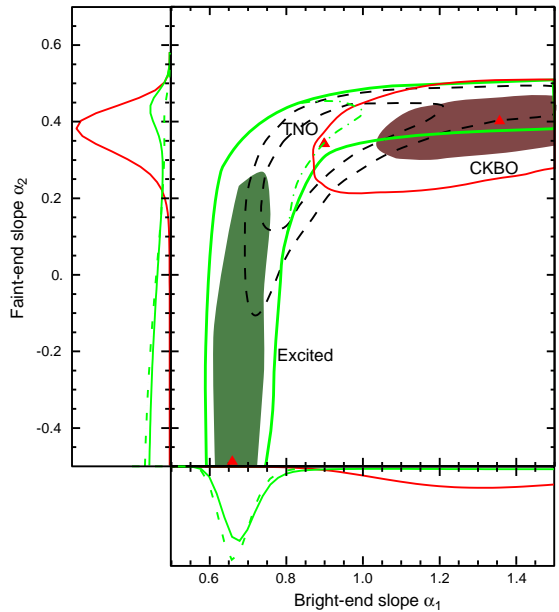}
\caption[]{\small
Allowed ranges of the two slopes for double-power-law fits to the
differential surface density of TNOs, as per \eqq{plaw2}.  
Shaded regions enclose 68\% of posterior probability for the CKBO and
Excited subsamples (red and green 
tint respectively), with outer solid contours bounding 95\% regions.
The dashed contours are for
the full TNO sample. 
Along the horizontal and vertical axes are the projected 1-dimensional
distributions of each slope.
The two dynamical classes have distinct magnitude distributions, with
the exception of the high-$\alpha_1$ tail on the outer Excited
contour.  If we include a prior constraint that the Excited class
contain one object on the sky with $R\le18.5$ (``Quaoar prior''), we
obtain the dash-dot contours instead.
The
bright-end slope of the CKBO group is likely steeper than the Excited
class, and the faint-end slope 
of the Excited class is probably shallower.
}
\label{a1a2contour}
\end{figure}

In Figure~\ref{a1a2contour} we plot the Bayesian posterior
distribution $P(\alpha_1,\alpha_2)$ for the double-power-law fits to
the various samples after marginalization over the less interesting
variables $\Sigma_{23}$ and $R_{\rm eq}$.  We also plot the
projections onto the single variables $\alpha_1$ and $\alpha_2$ for
the CKBO and Excited subsamples.  Note that we have applied a prior
restriction $-0.5<\alpha_i<1.5$ as we consider the more extreme slopes
to be unphysical.

Several features of the $\alpha_1,\alpha_2$ constraints are
noteworthy.  First, the CKBO and Excited samples once again appear to
be distinct, except that the 95\% CL region for the Excited sample has
a tail at $\alpha_1>0.8$, $\alpha_2\approx0.4$, that contains 10\% of
the posterior density, and overlaps the CKBO region of viability.
Fits in this secondary range predict very few Excited TNOs at $R<19$,
which is consistent with our limited sample, but inconsistent with the
membership of Pluto, Quaoar, and/or 2004~DW in the Excited class.  If
we include in our likelihood function a prior equal to the probability
that each model 
produce at least 1 Excited TNO at $R\le18.5$ (the ``Quaoar prior''),
then this long tail 
disappears from the Excited credible region (as illustrated by the
dash-dot lines in Figure~\ref{a1a2contour}).  We further discuss the
CKBO/Excited dichotomy in \S\ref{twosamples}.  

For any $\alpha_2<0.6$, the mass integral converges at
the faint end (\S\ref{massintegral}), and this is satisfied at high
confidence for all samples.  The faint-end slope of the CKBOs is well
constrained at $0.38\pm0.12$ (95\% CL).  The faint-end slope of the
Excited class is poorly determined, with only a bound
$\alpha_2\lesssim0.36$ (95\% CL with the Quaoar prior).  The absence of Excited
TNOs in the ACS survey leads to this degeneracy.

For $\alpha_1>0.6$ the bright-end mass
converges, and this is satisfied at 95\% confidence for the Excited
subclass and at very high confidence for the CKBO sample.  The
bright-end slope for the Excited class is $0.66^{+0.14}_{-0.08}$ (95\%
CL with Quaoar prior),
while the absence of bright CKBOs leads only to a bound of
$\alpha_1\gtrsim0.97$ for their asymptotic index.

The CKBOs certainly seem to have a steeper bright-end
slope (fewer large objects) than the Excited objects, 
and there is less secure evidence that the Excited class has a shallower
faint-end slope (fewer small objects).  There is thus 
evidence for different accretion and erosion histories for these two
samples. 

\subsection{Internal Consistency}
\label{consistency}
Before proceeding with further interpretation, we pause to ask whether
there are any internal inconsistencies among the collected survey
data. 
We check the surveys individually for consistency with the best-fit
double power law.  

We will make use of three consistency tests.
The first is simply the number $N$ of detected objects.
This probability distribution for $N$ follows the Poisson
distribution,
\begin{equation}
P(N|\bar N) = {\bar N^N \over N!} e^{-\bar N}.
\label{poisson}
\end{equation}
and the cumulative probability
$P(\le N)$ of having detected $N$ or fewer objects is also easily
calculated.  The drawback of the Poisson test is that it makes no use
of the distribution of magnitudes within a survey.

The second statistic that we use is $Q_{AD}$ statistic described above,
which has the
disadvantage that it discards information on the total number of
detected TNOs.

The third statistic we will use is a form of log-likelihood:
\begin{eqnarray}
\label{logl}
\logl(\{m_i\} | g) & \equiv & N \log \bar N - \log N! - \bar N
	\nonumber \\
 & & + \sum_{i=1}^N \log [g(m_i)/\bar g] \\
\log \bar g & \equiv & {\int g \log g\, dm  \over  \int g(m)\,dm}
\end{eqnarray}
This statistic is useful in that it is the log of $L$ in \eqq{gladman} when
the model and number of detections $N$ are fixed.  The expectation value of
$\logl$ when $N$ is held fixed is also equal to the log of
the Poisson probability in \eqq{poisson}.  This statistic hence has
sensitivity to both the number and distribution of detections in a
survey under test.  For any given model and survey, we can generate
1000 or more Monte-Carlo realizations to calculate the probability
$P(<\logl)$ of 
the measured likelihood being generated by chance under the model.

When comparing a constituent survey to the full-data fits, we do not
re-fit each realization, because individual surveys do not heavily
influence the overall fit.

The results of the three statistical tests for each
survey are given in Table~\ref{surveys}.  The only sources of tension,
marked in boldface in the Table,
are for the Larsen \etal\ survey, which
contains too many objects at 97\% CL, and the Gladman \etal\ survey,
which is overabundant at 98\% CL.  The AD tests also indicates that
the Larsen \etal\ survey is too skewed toward faint objects
($Q_{AD}=0.05$) and the Gladman \etal\ detections
are also too skewed toward faint objects ($Q_{AD}=0.03$).
Chiang \& Brown were also slightly lucky to find 2 objects in their 0.01
deg$^2$ survey, if the collective fit is correct.

These excursions are worse than we would expect from Poisson
statistics, but not horribly so: with 3 statistical tests for each of
7 surveys, we expect $\approx 1$ to show a discrepancy at $>95\%$
significance, while we have two surveys discrepant at this level.
There is no justification for
excluding any particular survey data.  For example, consider 
the fact that the ABM and Gladman \etal\ surveys are in poor
agreement in the $25<R<26$ magnitude range. The Gl sky density in this
bin is 3 times that of ABM.  The odds of obtaining by chance a
disparity this large given the number of detections in the survey are
approximately 1\%.  Since we have overlap between different surveys in
several of our bins, the chance of our having found one such
discrepancy between any two surveys in any of our bins is perhaps 5\%.
The discrepancy is hence worrisome but not outrageous.

It is possible
that either ABM overestimate their completeness or Gladman \etal\ have
some false-positive detections at their faint end.  There are no
obvious flaws to either work---ABM have a thorough artificial-object
estimate of efficiency, and Gladman \etal\ detect each object on two
consecutive nights.  In the absence of any reason to reject either
dataset, we will continue to sum the effective areas and total
detections of both surveys.  We have verified that none of our
conclusions are significantly affected by omission of either dataset.

Further data in this magnitude range are clearly desirable, as it
helps define the departure of the faint end from a power-law slope.
Surveys at $25<R<26$
require long integrations on large telescopes with large-area CCD
mosaics.  Such efforts are underway, using for example the VLT
(O. Hainaut, private communication)
and Subaru (D. Kinoshita, private communication) telescopes.

\subsection{Dynamical Subclasses}
\label{twosamples}
The parametric $\Sigma(R)$ fits to the CKBO and Excited 
subclasses appear to differ, though the evidence is not yet ironclad.
Since the CKBO and Excited subclasses have the same effective area at
a given
magnitude, we may apply a 2-sample Anderson-Darling test to see if
their magnitudes are drawn from the same distribution.  We obtain
$Q_{AD}=0.039$, {\em rejecting at 96\% confidence the hypothesis that
the CKBO and Excited TNOs have identical magnitude distributions.}
The largest difference in the
magnitude distributions is the lack of bright CKBO members, which is
noted by TB and discussed in detail below.  Nearly all the statistical
significance of the result arises from the TJL sample.
The test indicates that
there are (at least) two distinct size distributions in the TNO
population, and hence the magnitude distribution should be fit by
dynamical class rather than summed.  

Our division into dynamical classes is crude because of incomplete
orbital elements (\S\ref{compendium}), 
which can only have ameliorated the distinction
between the two size distributions.  A more precise division may yield
even more pronounced size differences between dynamical classes.

\section{Interpretation}

\subsection{The Mass Budget of the Kuiper Belt}
\label{massintegral}
The detection of departures from a single power law now make it possible
to estimate the total TNO mass without any divergences.  The total
mass of a TNO population may be expressed as
\begin{eqnarray}
M_{\rm tot} & = & \sum_{\rm TNOs} M_i \\
 & = & M_{23} \Omega \int dR\, \Sigma(R) 10^{-0.6(R-23)}
\,f^{-1} \nonumber \\ 
& & \left\langle 
\left( {p \over 0.04} \right)^{-3/2}
\left( {d \over 42\,{\rm AU}} \right)^6
\left( {\rho \over 1000\,{\rm kg}\,{\rm m}^{-3}} \right)
\right\rangle.
\label{totmass} \\
M_{23} & = & 7.8\times10^{18}\,{\rm kg}.
\label{m23}
\end{eqnarray}
The surface density $\Sigma$ is the mean over solid angle $\Omega$ of
the sky, and $f$ is the fraction of the TNO sample at magnitude $R$
that lies within the area $\Omega$.  The material density, albedo, and
heliocentric distance are $\rho$, $p$, and $d$, and $M_{23}$ is the
mass of a TNO that has $R=23$ with the given canonical albedo,
density, and distance.  We ignore the effects of illumination phase,
heliocentric vs geocentric distance, and asphericity.  The
angle brackets indicate an average over the TNOs at the given
magnitude.  
We make the usual bold assumption that the bracketed quantity is
independent of apparent magnitude and can hence be brought outside the
integral in \eqq{totmass}.  We will carry the integral from
$14<R<31$.  

\subsubsection{The Mass of the Classical Kuiper Belt}
The approximation of a common heliocentric distance is workable for
the CKBO sample, which by definition ranges from $38<d<55$~AU.  Of the
nearly 1000 TNOs detected to date, none are known to have
low-inclination, low-eccentricity orbits with semi-major axis
$a>50$~AU or $a<38$~AU. A sharp decrease in surface density beyond 55~AU
is apparent even after correction for selection effects
\citep{ABM01,TJL,TB01}. It is therefore physically meaningful to consider
our CKBO sample to represent a dynamical class that is largely
confined to heliocentric distances of $42\pm10\%$~AU. 

For the CKBO sample, the value of the integral is
$(2.85\pm15\%)\,{\rm deg}^{-2}$ (95\% CL)
when marginalized over the double-power-law fits.  The value 
when using the rolling-index form for $\Sigma(R)$ is
indistinguishable, so we believe this to be robust to
parameterization.  The solid angle over which $\Sigma$ has been
averaged is $360\arcdeg\times 6\arcdeg$.  Because the CKBO sample is
by definition restricted to $i<5\arcdeg$, the residence fraction $f$
is high---unity for $i<3\arcdeg$, and 0.83 if the inclinations are
uniformly distributed between 0\arcdeg\ and 5\arcdeg.  We will take
$f=0.9$ and ignore any uncertainty as it will be small in comparison
to that of the albedo and density terms.  We then obtain (at 95\% CL)
\begin{eqnarray}
\label{mckb}
M_{\rm CKB} & = & (5.3\pm0.9)\times 10^{22} \, {\rm kg} \nonumber \\
 & & 
\left( {p \over 0.04} \right)^{-3/2}
\left( {d \over 42\,{\rm AU}} \right)^6
\left( {\rho \over 1000\,{\rm kg}\,{\rm m}^{-3}} \right)
\end{eqnarray}
The prefactor is now determined to much greater accuracy than the
scaling constants.  The CKBO mass is nominally equal to just
$0.010M_\oplus$ or a mere 4 times the mass of Pluto (8 times, if the
CKBOs share Pluto's density of 2000~kg~m$^{-3}$).

TJL report an estimated mass for CKBOs with diameters in the
range $100\,{\rm km}<D<2000\,{\rm km}\approx1.8\times10^{23}\,{\rm
kg}$, under the same assumptions about density and albedo as made
here.  We would naively expect our estimate to be larger, not three
times smaller, than the TJL estimate, because we now include bodies
smaller than 100~km---though the ACS data show that these smaller
bodies hold a minority of the mass.  The discrepancy is in part
attributable to our more restrictive definition of the classical belt:
we require $i<5\arcdeg$ whereas TJL demand only $41<a<46$~AU and
$e<0.25$.  The larger part of the discrepancy is due to to our
conclusion that the CKBO population has $\alpha>0.6$ for
$m\lesssim24$, greatly reducing the mass that TJL's $\alpha\approx0.6$
places in large objects.

\citet{Gl01} estimate the mass of TNOs in the 30--50~AU range to be
0.04--0.1~$M_\oplus$ (for unspecified material density), if the size
distribution turns over to the Dohnanyi slope for $D\sim40$~km.  Our
new value remains below this estimate even if we allow for the fact
that the Excited TNOs contribute a similar mass density to the CKBOs
in this distance range (next section), again reflecting the fact that
the present data fall below the assumed power laws at both large and
small object sizes.

\subsubsection{The Mass in Excited TNOs}
The TNOs in our Excited sample---a mixture of resonant and
high-excitation non-resonant orbits---are drawn from dynamical
families whose radial and vertical extent remains quite uncertain.  In
particular our selection effects for the so-called scattered-disk
objects are poorly known.  Our estimates of the total mass 
will therefore be much less secure than for the CKBO population.

The assumption that the bracketed quantity in
\ref{totmass} is independent of magnitude is dubious for the
Excited class, because it is likely that the brighter bins are biased
toward objects near perihelion, which can be an extreme bias for a
scattered-disk member.  We proceed nonetheless.  The integral
over the double-power-law $\Sigma(R)$ is well determined:
marginalizing over the fitted parameters we obtain 
$(3.5\pm1)\,{\rm deg}^{-2}$ for the double power law, somewhat
lower for the rolling-index fit.

Because the excited population has high inclinations, the residence
fraction $f$ will be smaller; if the inclinations $i$ are distributed
as a Gaussian with $\sigma_i=10\arcdeg$, then they spend on average
$f\approx0.5$ of their time within 3\arcdeg\ of the invariable plane.
The inclination distribution is, however, poorly known, especially
under our definition of the Excited sample.

The mean distance appears as $d^6$ and is highly uncertain for the
``scattered-disk'' objects, but is sensibly bounded for Plutinos.  
An upper bound on the Plutino mass comes by
assuming that the Excited class contains most of the Plutinos that are
closer than 39~AU to the Sun, or very crudely half of the total Plutino
population.
The brighter surveys are biased against
Plutinos that are currently beyond their semi-major axis distance, 
and under our definition of the CKBO sample, Plutinos at low
inclination beyond 38~AU will be put into the CKBO class.  So we 
will calculate a total mass for the detectable Excited sample assuming
$d=39$~AU, and
double it to bound the Plutinos.  Following this procedure, we obtain
\begin{eqnarray}
\label{mplutino}
M_{\rm Plutino} & \lesssim & 1.3\times 10^{23} \, {\rm kg}
\left( {f \over 0.5} \right)^{-1} \times \nonumber \\
& & \left( {p \over 0.04} \right)^{-3/2}
\left( {d \over 39\,{\rm AU}} \right)^6
\left( {\rho \over 1000\,{\rm kg}\,{\rm m}^{-3}} \right).
\end{eqnarray}

An estimate for the so-called ``scattered-disk'' population is even
less certain given the potentially large values of $\langle d^6\rangle$ 
(even though no detected objects\footnote{
We do not include the newly-discovered object Sedna at $d=90$~AU in
the scattered disk given its perihelion at 76~AU.} yet exceed $d\sim60$~AU).
Taking the
mean $d$ to be 42~AU and $f=0.5$ gives an Excited-class mass of
$1.3\times 10^{23}$~kg.
It perhaps suffices to estimate that the Excited sample is 
comparable to or several times the mass of the CKBOs, because the
surface-density integrals are nearly equal, the residence fraction $f$
is lower for the Excited objects, and the mean $d^6$ factor could be
larger for the Excited class.  The value here is, however, a
fair estimate of the Excited mass {\em within 50 AU.}

It thus appears that Pluto itself accounts for $\gtrsim10\%$ of the
mass of the Plutino population, and perhaps of the entire Excited
population.  We will investigate this further in \S\ref{brightest}.

TJL crudely estimate the number of $D>100$~km objects with
scattered-disk orbits to be comparable to the CKBOs, and the Plutinos
to be $\sim20$ times less abundant.  The absence of Plutinos in the
ACS data suggests that the Plutinos also have a break in $\Sigma(R)$
such that little of their collective mass is in small objects, 
but we do not attempt a further quantitative bound.  If this TJL
estimate is correct, then Pluto dominates the Plutino mass and is
several to tens of percent of the scattered-disk mass.

\subsection{The Largest Objects in the Dynamical Classes}
\label{brightest}
We examine here some implications of the large-size behavior of our
fitted $\Sigma(R)$ functions.  These results are not of course a
direct consequence of the new small-end ACS data, except insofar as
the latter make it clear that single-power-law fits to the full
population should not be expected to properly characterize the bright
end.  The results of this section flow from our efforts to create a
homogeneous subsample of all the published brighter surveys.

The double-power-law fits to the CKBO
sample favor a bright-end slope that is significantly steeper 
than for the Excited sample.
If correct, this implies a different accretion history for the two
classes, since it is believed that erosion should not have affected
the largest TNOs \citep{S95}.  It would also imply that the brightest
(and largest) member of the Excited class will be significantly larger
than the largest CKBO, and that the Excited sample will have a
substantial fraction of its mass in the largest objects (since
$\alpha_1\approx 0.6$) while the CKB will not.
A possible preference for large objects to reside in
high-excitation populations has been suggested before in studies of less
well-controlled samples \citep{LS01} and in theoretical studies
\citep{Go03}, and is verified by the TB survey.

We have inferred the bright-end slope difference by using only the
subset of survey data to date that has invariable latitude
$<3\arcdeg$ and has published survey statistics.  We may check the
accuracy of our inferences by some comparisons with the full sample of
nearly 1000 TNOs that have been discovered to date, as listed on the
Minor Planet Center web pages July~2003.

We first ask what should be the brightest TNO on the full sky if
extrapolation of our fitted double-power-law $\Sigma(R)$ were to
describe correctly all TNOs.  We may calculate the probability
distribution for the magnitude of the brightest TNO by marginalizing
over all our double-power-law fits, and assuming an available area of sky
equal to $\Omega/f$, using the values estimated above.
To be specific, we calculate
\begin{eqnarray}
P[ N(<R)=0 ] & = & \int d{\bf p} \,P({\bf p} |
 \{m_i\})\, P[ N(<R)=0 | {\bf p}] \\
 & = & \int d{\bf p} \, P({\bf p} | \{m_i\}) \nonumber \\
 & & 
    \exp\left[-(\Omega/f) \int_0^{m^\prime} dm^\prime\, 
	\Sigma(m^\prime|{\bf p})\right].
\label{brightformula}
\end{eqnarray}
Here ${\bf p}$ is a vector of parameters for $\Sigma$, $P({\bf p} |
\{m_i\})$ is the normalized Bayesian posterior probability for the
model parameters given the observed data, and $\Omega=360\arcdeg\times
6\arcdeg$ is the total area of the low-invariable-latitude strip on
the sky.

\begin{figure}
\plotone{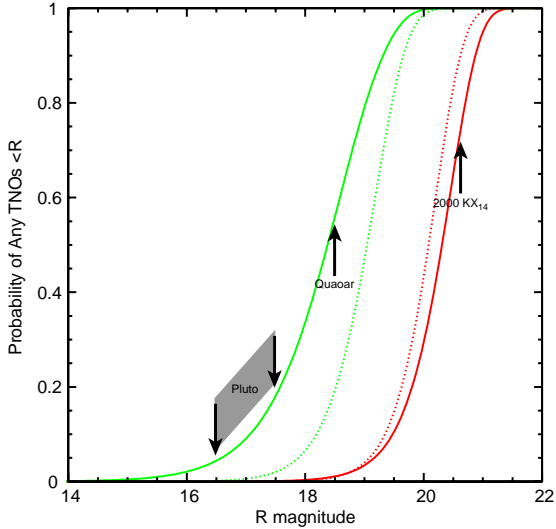}
\caption[]{\small
The distribution for the brightest expected TNO on the sky is shown vs
$R$ magnitude for the CKBO (red, right) and Excited (green, left) samples.  The
plotted quantity is the cumulative likelihood of finding a TNO
brighter than $R$.  The solid lines are derived from an
extrapolation of the double-power-law fits to the invariable-plane
surveys collected in this paper, and the dashed lines are for the
rolling-index power laws.  Given that much of the relevant sky area
has now been searched for bright TNOs, we see that the extrapolations
of the power laws
are consistent with 2000~KX$_{14}$ being the brightest CKBO (or nearly
so), and with Pluto or Quaoar being the brightest Excited TNO [Pluto
is plotted at the apparent magnitudes it would have with albedo 0.04--0.10.]
If the rolling-index fit is the proper extrapolation, Pluto would
have to be considered an exceptional object.
}
\label{brightplot}
\end{figure}

Figure~\ref{brightplot} shows
our estimated probability of there being, somewhere on the sky, a TNO
brighter than a
given $R$ magnitude.  For the CKBOs, the double- and rolling-power-law
fits are in general agreement.  The median expected magnitude for the
brightest CKBO on the sky is 20.3~mag.
The brightest TNO found to date that meets our CKBO
criteria is 2002~KX$_{14}$, at $R=20.6$, which is found by the TB
survey but was not used in our analysis since it is below our adopted
20.2~mag cutoff.  Nearly all the sky within 5\arcdeg\ of the
invariable plane has been surveyed for such bright objects, by TB
and/or the Southern Edgeworth-Kuiper Survey \citep{SEKS},
with all discoveries having been transmitted to the
MPC.  Hence it is unlikely that a CKBO significantly brighter or
larger than  2002~KX$_{14}$ exists on the sky.
The fitted and extrapolated $\Sigma(R)$ models
suggest a 70\% chance of finding a CKBO brighter than 2002~KX$_{14}$,
so it would be acceptable for 2002~KX$_{14}$ truly to be the 
largest CKBO (or nearly so).

For the Excited class, an extrapolation is more speculative since a
smaller fraction of the available sky has been surveyed and we include
in our analysis only a fraction of the surveyed area.  The median
expected magnitude of the brightest Excited TNO is 18.4 for the double
power law, and 19.0 for the rolling-index fits.  
The brightest TNO is Pluto, which would have $R=17.5$ if it had our
assumed albedo of 0.04 \citep{Bu03}, or $R=16.5$ if the albedo is
$0.10$, comparable to Quaoar \citep{Quaoar}.
Under the double-power-law fits, there is
4--17\% chance of finding an Excited TNO this bright, so it is
feasible to consider Pluto to have been produced by the same physical
process as the other TNOs.  Under the rolling-index model, however,
the odds of a Pluto on the sky are $<2\%$, hence under this model
we would have to consider Pluto as an exceptional object with a
distinct formation mechanism from the other TNOs.
The largest and brightest TNO outside the
Pluto/Charon system is Quaoar at $R=18.5$---also discovered by TB but
outside the bounds of our $\pm3\arcdeg$ swath---which at $i=8\arcdeg$ would
be excluded from our CKBO sample.  From the Figure we see that if the
bright end follows a power law, we would expect the largest Excited
TNO to be in the Quaoar--Pluto range.  So we again
find that the largest objects known to date are consistent with extrapolation
of our double-power-law $\Sigma(R)$ fits---or with the rolling-index
model, if we consider the Pluto-Charon system as exceptional.  

The degree of dominance of the brightest known objects is also
consistent with our double-power-law fits to our selected subsample.
The ten brightest known CKBOs have $20.6<R<21.8$ and contribute
$\approx2\%$ of the estimated CKB mass (assuming common albedo and
material density).  Again, surveys for
low-inclination TNOs this bright are now majority complete, so this
fraction is not likely to evolve much.  Hence the largest objects in
the CKB do {\em not} hold a significant fraction of its mass,
confirming the validity of the  $\alpha_1>0.85$ bright-end slope.

The ten brightest known resonant/scattered TNOs, excluding Pluto and
Charon, have $18.5<R\lesssim20.5$.  Their combined mass, with
canonical albedo and density, is $1.3\times10^{22}$~kg, roughly equal
to Pluto's mass, and roughly 10\% of the total estimated Excited
family mass.  Thus the largest bodies known hold tens of percent of
the known Excited-class mass, confirming that the power-law slope 
is near $0.6$.

The conclusions we have drawn from our controlled TNO
subsample are therefore supported by analysis of total sample of known
bright TNOs. In particular we find that the Excited class is near the
$\alpha_1=0.6$ value with equal mass per logarithmic size bin, while
the CKBOs have steeper $\alpha_1>0.85$ that puts less mass in large
objects.  Under the double-power-law model, Pluto is a uniquely but not
anomalously large Excited TNO.  Since erosion should not have
significantly altered the $\gg100$~km objects in the current
trans-Neptunian environment, a simplistic interpretation is that the
Excited objects were formed in a region of the Solar System that
allowed the accretion process to proceed further than in the
$\sim42$~AU location of the CKBs.  This is consistent with the
scenario that the Excited TNOs formed in a low-excitation
population closer to the Sun \citep{Ma93,Ma95,Go03}, where higher space
densities
and relative velocities could speed accretion.

\subsection{The Smallest Known TNOs and Erosion History}
Calculations of collision rates and energies suggest that the
population of TNOs with $D\lesssim100$~km has been strongly influenced
by collisional erosion.  The ACS survey provides the first strong
constraints on this population, so in this section we provide some
basic comparison of the observed magnitude distribution to models of
the accretion/erosion history of the TNOs.  Throughout this section we
will make the simplifying assumption that the TNOs samples we have
gathered are at fixed distance with fixed albedo, so that the
magnitude distribution may be directly translated into a size
distribution.  If the magnitude distribution
scales locally as $dN/dm\propto 10^{\alpha m}$, then the
size distributions will locally follow the common parameterization
$dN/dD\propto D^{-q}$ with $q=5\alpha + 1$.  
We will prefer the double-power-law models for $\Sigma(R)$ to the
rolling-index models for this discussion, as the size-evolution models
typically predict such behavior.

\subsubsection{Comparison with the MPC Database}
The conclusion that the faint-end $\Sigma(R)$ is shallower for
CKBs than Excited objects does not have exceptionally strong
statistical support in our sample, {\it e.g.} 3 vs 0 detections in the
ACS survey.  We again look to see if the trend is
borne out by the sample of all objects submitted to the MPC.  Of the
23 objects with $H\ge9.6$ and perihelion $q>28$~AU, only 5 would fall
into our Excited class.  The dominance of CKBOs in this range is
significant given that: Excited TNOs dominate the low-$H$ objects are
comparable to CKBOs near $R\approx23$; discovery of high-$H$ Excited objects
is favored because of their nearer perihelia; and several of the
$H>9.6$ Excited TNOs could be CKBOs once more accurate inclinations
are obtained.  This test is not independent of our more controlled
analysis because many of the $H>9.6$ TNOs at the MPC are from the ACS,
ABM, and Gl surveys.  Nonetheless the trend in our controlled
subsample seems to extend to the full sample.

\subsubsection{Equilibrium Collisional Cascades}
An analytic description of a collisional cascade is given by
\citet{Do69}, who derives $q=3.52$, $\alpha=0.50$, for an equilibrium
solution.  In such a steady state, the loss of objects from a given 
mass bin due to grinding and catastrophic destruction is exactly 
replenished by the gain of fragments from collisions on larger
bodies.  A population of high-mass objects is of course required as a
mass reservoir to maintain such an equilibrium.  We see from
Figure~\ref{a1a2contour} that $\alpha_2\ge0.5$ has a low probability
of describing either the CKBO or Excited populations in the ACS
survey:  the true distribution is markedly shallower (meaning
relatively fewer small bodies) than the canonical Dohnanyi value,
particularly in the Excited sample.

The failure of the analytical model could perhaps be ascribed to the
particular assumptions in the model, {\it e.g.}, that the size
distribution will follow a power law and that the distribution of
fragment sizes will take a particular power-law form, although
the equilibrium form should be robust as long as the fragmentation law
is scale-free.  \citet{PS04} suggest that the observed size
distribution can be explained by a fragmentation law for rubble-pile
objects which is not scale free.
An
alternative explanation is that the present-day TNO population is not
in collisional steady state. Perhaps we have instead a snapshot of two
populations
for which the small-body population is continuously decreasing and
erosive destruction is very advanced.  In the CKB, it is not even
clear what bodies could serve as a mass reservoir, since the largest
bodies hold little of the mass.

\subsubsection{Numerical Models: General Results}
Modelling of non-equlibrium collisional evolution requires numerical
simulation, particularly if coeval accretion is modelled as well.
Three groups have produced such models: \citet{DS00} (who estimate
body lifetimes but do not follow the long-term evolution of the size
distribution), \citet{KL99}, and \citet{DFW}, and the antecedents of
these papers, particularly \citet{S95} and \citet{DF97}, offer results
of numerical simulations of TNO size distributions.  A review is given
by \citet{FDS}.

Before comparing the numerical models to the new data in detail, we
first review some of the general conclusions of these works, and
re-examine them in light of the new observational results, namely: the
detection of a strong break to shallower distributions at
$D\lesssim100$~km, implying much lower impact rates;
the stronger bounds on the bright-end $\alpha$; and
the clear detection of size-distribution differences between two
dynamical populations.

A simple analytical result of \citet{S95} is that for a population
with mean eccentricity $0.03\lesssim \langle e \rangle \lesssim 0.1$
at $\approx42$~AU, collisions are on average erosive for objects
smaller than a critical diameter $D^\star$ and accretional for larger
bodies.  The critical diameter $D^\star$ is 100-300~km for the
``strong'' bodies of \citet{DS00}, which appear at $R$ magnitudes
23--25.5 under the canonical albedo.  Note from Figure~\ref{fitfig}
that this is the magnitude range where $\Sigma(R)$ develops a
significant deficit relative to the bright-end power law.  {\em The
TNOs that are ``missing'' are those that are susceptible to
collisional destruction.}

A general conclusion of \citet{S95} and later models is that the
collision rate in the present trans-Neptunian region is too low for
$D\gtrsim200$~km objects to have been formed by pairwise accretion in
4~Gyr.  The ACS results only exacerbate this difficulty since the mass
in small bodies is found to be much lower than all the models had
assumed.  We note, however, that the maximum size of bodies in the CKB
sample is significantly lower than that in the Excited sample, with
the largest known CKBO being 60 times less massive than Pluto.  These
largest
bodies should be largely unaffected by erosion according to all
models.  We can conclude that the Excited population is ``older'' than
the CKB in the sense of cumulative number of accreting collisions.  
This is consistent with an origin for the Excited bodies a smaller
heliocentric distances, in a denser and faster-moving section of the
disk than the CKB.

\citet{DS00} calculate impact and disruption rates for TNOs in the
present Kuiper Belt, concluding that all TNOs should be heavily
cratered but that the lifetime against catastrophic destruction for
$D\gtrsim1$~km bodies exceeds the age of the Solar System.  We now
believe the space density of $D<100$~km bodies to be much lower than
assumed in these models, perhaps by several orders of magnitude for
km-scale bodies.  A recalculation of present-day impact rates could
show that small (km-scale) TNOs 
have negligible cratering in the past billion years, and even
large TNOs would have recent craters only from sub-km impactors.  This
would argue against a collisional resurfacing effect as the source of color
diversity among TNOs.  

So we find the apparent paradox that objects small enough to be
subject to collisional disruption are strongly depleted, yet have
lifetimes against disruption that are longer than the age of the Solar
System.  A possible resolution of this paradox is that we now see the
end state of the erosion process:  the original CKB (and perhaps the
precursor region of the Excited population) was more massive and
richer in small bodies, leading to strong erosion and depletion.  The
erosion lowered the volume density to the point where lifetimes
exceeded $10^9$ years, leaving the present ``frozen'' population
behind.   In this scenario, present-day $D\lesssim100$~km bodies are
likely all 
collision fragments, but they may have had little alteration in the
past 3 billion years.  Surfaces would be heavily cratered but old.

At some point in the history of the CKB, the Excited population
started to cross the CKB, with higher collision velocities, which
would have exacerbated the erosive depletion process.

Alternatively, the accretion process in the original source disk was
extremely efficient at collecting 1--10~km bodies into
$D\gtrsim100$~km bodies---but not at producing 1000~km bodies from these.  A
quantitative explanation for the preferred scale of accretion would be
required. 

\subsubsection{Comparison with Detailed Models}
\citet{KL99} and \citet{DFW} plot the time history of the TNO size
distribution in their numerical accretion/erosion models.  The
following seem to be generic features of the models:  at the large
end, the distribution becomes shallower with time as the size of the
largest accreted objects increases.  At the small end, the slope
either approaches the Dohnanyi value \citep{KL99} or becomes
progressively shallower \citep{DFW}.  

The observed CKB and Excited distributions disagree with the
\citet{KL99} models in several respects:  first, the observed faint-end slope
does not match the Dohnanyi value they predict; the models predict a
large excess of bodies at the transition region, which is not
observed; and the observed transition appears to be at the 10-100~km scale, not
the km~scale predicted by the models.  These models, however, did not
attempt to model the effect of several Gyr of further collisional
erosion after the formation epoch.

The agreement of the data with the \citet{DFW} models is fairly
strong.  The last time slice (1~Gyr) of their simulation shows bright- and
faint-end slopes of $\alpha\approx0.6$ and $\approx 0.0$,
respectively, with a transition near $D\approx50$~km.  This is
remarkably similar to the observed Excited sample magnitude
distribution.  

The CKB sample resembles younger time slices in these models
($\sim200$~Myr) in having steeper slopes at the bright end
($\alpha_1\approx1.2$) and faint end ($\alpha_2\approx0.4$), as well
as a smaller maximum object size.  The transition region is at smaller
diameters in the models than in the data, however.  It appears in any
case that the CKB is less evolved than the Excited sample in an erosion
sense as well as an accretion sense.  This might be expected from the
smaller velocity dispersion of the present-day populations (if erosion
postdates any migration of the Excited bodies) or from the higher
space density and/or velocity of the Excited source region (if erosion
predates migration).

The ACS data should serve as a target for future size evolution
models.  It may be particularly important for future models to
consider the coupled evolution of the two (or more) co-spatial
dynamical populations that appear to exist currently and to have
different size distributions.  An interesting question is whether {\it
any} dynamical mechanism is required to reduce the mass of the CKB to
its present $\approx0.01 M_\oplus$ value from the $\approx 10 M_\oplus$
levels that appear necessary to support large-object accretion, or
whether erosive removal is sufficient.  This question is particularly
interesting in light of the absence of detected TNOs beyond 55~AU, as
discussed below.

\subsection{The Source of the Jupiter-Family Comets}
It is currently widely accepted that the Jupiter family short period 
comets (JFCs) and the Centaurs are objects that have escaped from the
Kuiper Belt. Our HST-ACS survey provides an estimate of the population of
cometary precursors in the Kuiper Belt, which we can now compare
with estimates obtained from dynamical models of the Kuiper Belt-JFC
connection.  
In one of the earliest modern models of the Kuiper Belt--JFC connection,
\citet{HW93} estimate that a population of $5\times10^9$
cometary precursors in the Kuiper Belt in the 30--50 AU range is required
to account for the observed population of JFCs.  
A more detailed calculation of essentially the same physical model by
\citet{LD97} revises this estimate to $7\times10^9$.
These calculations are based on what can now be described as a 
``cold'' Kuiper belt, with most KBOs in low eccentricity, low inclination 
orbits. In another model, \citet{DL97} postulate the ``scattered
disk'' as a source of JFCs, and estimate a population of $6\times10^8$ 
cometary precursors in the entire scattered disk, of which $\sim1.4\times10^8$
are at heliocentric distances 30--50 AU. 
In yet another calculation, \citet{Mo97} models the Plutinos as a 
source of the JFCs, and estimates a required source population of
$4.5\times10^8$ Plutino comets.  A commonality of these models are the
assumptions that the various  
classes of the Kuiper Belt constitute stable reservoirs of cometary precursors,
and that some fraction of these objects escape into the inner solar system on 
gigayear timescales due to slow orbital chaos induced by the gravitational 
perturbations of the giant planets.
The models differ only in the initial conditions of the Kuiper Belt comets,
{\it i.e.}, in the choice of the KB subclass for the putative source of
the JFCs.

We can convert the modeled population estimates to a surface density 
by assuming that the projected sky area of these estimated populations is 
$10^4$ deg$^2$ (corresponding to a $\pm15^\circ$ latitudinal band around 
the ecliptic or invariable plane).  
A greater challenge is to define the magnitude range of cometary precursors
that correspond to the dynamical models.  CLSD took this to be
$R\sim28.5$, corresponding to an object of $\sim10$~km 
radius at heliocentric distance 40~AU, assuming the usual 4\% geometric 
albedo.  With this assumption, CLSD found agreement between their
measured surface density and the required cold precursor population.

\begin{figure}
\plotone{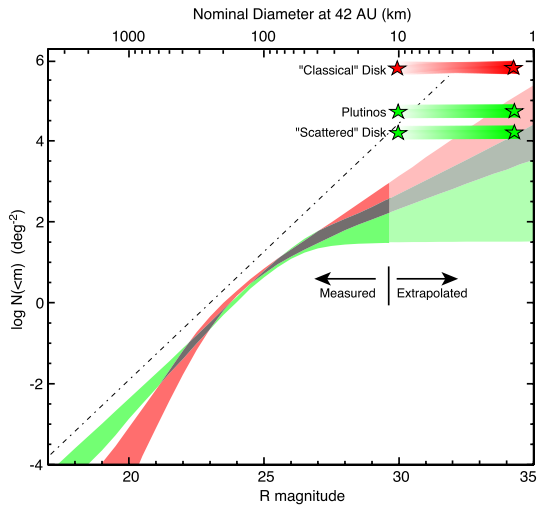}
\caption[]{
\small
The cumulative surface density of the Excited KBO population (green)
and of the classical population (red), parametrized by the double
power law (Equation~[\ref{plaw2}]).
The regions are
bounded by the 95\% confidence upper and lower bounds at each
magnitude.  Dot-dash line is the now-outdated power-law fit to the
total population from TJL.
The horizontal bands in the upper right are theoretical estimates
based on models  
of the Kuiper belt as a source of the JFCs; the assumed source
population for each case is labelled.  The horizontal extent is a
reminder that the precursor population of known JFCs is not currently well
defined, probably lying somewhere above 1~km diameter.
The observed populations fall short of the surface densities calculated
for putative JFC reservoirs unless a favorable extrapolation of the
Excited-class density is assumed.  The CKB population falls
$\sim10\times$ short of the required reservoir even with favorable
extrapolation. 
}
\label{jfcfig}
\end{figure}

We find a surface density three orders of magnitude lower
than the claim of CLSD.  The nuclei of JFCs have now, however, been 
measured to have diameters predominantly in the range 1--10~km
\citep{Lamy}.  
It is unfortunately unclear whether the JFC population is close to {\em
  complete} for $D\gtrsim1$~km.
It is unlikely that the JFC
  precursors would be smaller than the JFCs themselves, or that
  albedos could be significantly lower than 0.04, so we may consider
  $R\approx35$ to be the faintest possible precursor population.
The dynamical estimates of 
the surface density of trans-Neptunian cometary precursors are shown in 
Figure~\ref{jfcfig} by the horizontal bands in the upper right, which
indicate a range of 1--10~km as the required size of the true
precursors.  
Figure~\ref{jfcfig} also plots
the 95\% likelihood of the parameterized fits to observations, 
as described in \S\ref{constraints}.

Extrapolation of the ACS
measurements to fainter magnitudes is required to compare the TNO
densities to the putative required JFC source populations.
The classical belt and Plutino populations are insufficient to supply
the JFCs even under the most favorable extrapolations of the ACS
observations.  Of course we cannot rule out an upturn in the
$\Sigma(R)$ slope at $R>29$.  
A slope near $q=3.5$ ($\alpha=0.5$) for 1--20~km diameter 
Jupiter-crossing
bodies is suggested by \citet{Z03} based upon
crater counts on Europa and the bright terrain of Ganymede.  This
value is outside the allowed $\alpha_2$ range for the Excited TNOs and
marginal for the CKBOs.  The ecliptic-comet samples of \citet{Lamy}
show a size distribution with $\alpha=0.32\pm0.04$ or $0.38\pm0.06$,
depending upon the inclusion of NEOs with ``cometary'' orbits, in the
range $D>3.2$~km.  Either
range is marginally compatible with our Excited TNO sample, and agrees
well with the measured $\alpha_2$ for CKBOs.  
Of course many comet nuclei are known to undergo dramatic
transformation or disintegration upon entering the inner solar system,
so the match with TNO sizes may be fortuitous.  The apparent
discrepancy between our faint TNO size distribution and the crater
sizes on the Galilean satellites requires further investigation,
though it is interesting to note that both the TNO data and the crater
counts have size distributions that flatten to smaller sizes.

The scattered disk remains a viable precursor population, if the
precursors are near 1~km diameter and the slope at $R>29$ is the
steepest permitted for $R<29$.    Thus with a favorable choice of
precursor diameter {\em and} favorable extrapolation to $R\approx34$,
present scenarios for JFC origin in the scattered disk can work.

There are several factors of 2 (or more) still indeterminate in the
quantitative understanding of the origin of JFCs: the completeness of
the known population, the size and albedos of their precursors, the
dynamical delivery rate and lifetime of the JFCs, the extent of the
scattered-disk, and the extrapolation to fainter magnitudes.  
Any definitive judgement on the origin of JFCs would clearly be
premature at present.

\subsection{Constraints on the Distant Population}
The ACS survey fails to detect any object further than 42~AU despite
extending the TNO completeness limit by 2~mag from previous
work.  Several previous works \citep{ABM01, TJL, TB01} have shown that
the space density of objects of size $D\ge150$~km must decrease
significantly beyond 50~AU from the Sun, {\it i.e.}, the CKB is
bounded.  It is possible that there is an outer belt that is deficient
in objects of this size and hence undetected.  We provide here a crude
bound on an outer disk composed solely of smaller bodies based upon
the complete absence of distant objects at $R<28.7$ in the 0.02~deg$^2$ of
the ACS survey.

Take a simple model in which an outer belt is localized to
heliocentric distance $d$ and contains a population of objects of
diameter $D$.  We assume that the outer belt members have albedos and
material densities similar to the CKB, and that the projected mass density on
the invariable plane is some multiple $g$ times the value for the CKB
derived in \eqq{mckb}.  For $d=60$~AU, the magnitude limit for
the ACS survey corresponds to $D\ge37$~km (for $p=0.04$).  The
invariable-plane number density of an outer disk of such objects would
be $900g$ per square degree.  The ACS data limit the density at
95\% CL to be $\le150\,{\rm deg}^{-2}$, implying $g\le0.17$.  If $D$
of the outer belt is larger, the limit on $g$ is roughly unchanged,
because at brighter magnitudes the area surveyed to date is larger.
The rough conclusion to be drawn is that an outer belt must either
have most of its mass in objects with $D<40$~km, or must be substantially
less massive than the known CKB.  Caveats are that the presumed outer
belt must be near the invariable plane at the ACS field, must not be
more dispersed vertically than our definition of the CKB, and must not
have albedo or material density dramatically different from the CKB.

This limit on the mass of an outer Kuiper Belt does not apply to
objects with orbits like Sedna
which at $d=90$~AU is approaching its perihelion near
76~AU on a highly elliptical orbit \citep{Sedna}.  
While the ACS survey would easily detect the motion of bodies anywhere
on such orbits, they would be $>500$~AU distant most of the time, where even
objects 1000~km in diameter would fall below our flux
We also do not usefully
constrain the mass contained in rare large objects such as Sedna itself.

A dearth of CKB mass beyond $\sim$50~AU may be a result of the initial
conditions of Solar System formation.  \citet{W03}
attributes a lack of CKB mass to processes from the era of
planetesimal formation 4.5~billion years ago. In his model, solid
bodies of sizes 0.1--10~m drift inward from locations of
50--100~AU due to gas drag. Interior to 50~AU, sufficient
solid surface densities exist for accretion to continue. Large bodies
do not accumulate outside of $\sim$50~AU, and the exact location of
the outer edge depends on the initial disk density radial profile.
This model adequately produces the observed CKBO distribution and
implies that there should be little mass beyond 50~AU, even for bodies
with $D<40$~km.  Alternately, the outer region of the nascent Solar
System could have been eliminated entirely through photoevaporation of
the circumstellar disk due to UV~flux from nearby OB~stars ({\it e.g.},
\citet{Th01}).  Evaporation of the disk (and entrained grains)
occurs to radii as small as 50~AU in $10^5$~years in an Orion-type
environment.  The $10^8$~or $10^9$~year CKBO formation timescale would
therefore preclude formation of bodies outside the evaporation radius
of perhaps 50~AU. \citet{LM04} suggest that the full TNO population
accreted within $\approx30$~AU and was transported to its present
location by interaction with a migrating Neptune; in this case we
still need to explain the truncation of the original planetesimal
system at 30~AU.

The lack of CKBOs beyond 50~AU may instead be a result of dynamical
evolution of the outer Solar System subsequent to its formation.
\citet{AL01} considered the case in which the Solar System
had dynamical encounters with other members of its birth cluster;
these interactions could have disrupted the KB outside of 50~AU and
increased KBO eccentricities to~0.2 or larger. 
Later close stellar encounters could have had similar
effects: disrupting, exciting, or even removing the outer CKB
\citep{Ida00, LK01}.

The observed data is also consistent with the interpretation that
significant mass is present in the CKB outside of 50~AU and that this
mass resides entirely in bodies smaller than $\sim$40~km.  
It is conceivable that, within the standard model of accretion
of planetesimals in a (gaseous) circumstellar disk, accretion
timescales beyond 50~AU were simply too long for bodies larger than
40~km to form before the present day or some disruptive event.
We note that a substantial mass in $D>3$~km bodies in an outer Kuiper
Belt should be detectable by occultation surveys in the near future
\citep{TAOS}. 

A final possibility is that large bodies
formed in distant orbits but have since been largely eroded or
shattered to smaller sizes.  It is usually difficult to destroy the
largest bodies in a population, but the remaining ones could retain
a sufficiently small portion of the original nebular mass that they
remain undetected.

\section{Summary}

The superior coverage and efficiency of the Advanced Camera for Surveys
has allowed a survey for TNOs with $m\le29.2$ over 0.02~deg$^2$.  The
faint limit of the survey is determined by the criterion that a TNO
have signal-to-noise ratio $\nu>8.2$ over the 22~ksec total
integration time of the discovery epoch; false positive detections begin
to appear below this threshold.  A modest investment in present-day
CPU power allows us to track, in effect, the HST along each of
$\approx10^{14}$ possible TNO orbits.  

Three new objects are detected, compared to the $\approx 85$ that were
expected from extrapolation of the power-law fit to brighter TNO
survey detection rates.  Artificial TNOs implanted
into the data are recovered efficiently, verifying the completeness of
the detection process, and the reality of the three objects is
confirmed by their detection in a second set of ACS observations and
the detection of the brightest one ($R\approx27$) at the Keck
Observatory.

All three detected objects are consistent with nearly-circular,
low-inclination orbits near $a\approx42$~AU, although complete orbital
elements will require followup detection (with HST)\footnote{Cycle 13
observations on HST have been allocated for retrieval.}.
Assuming an
albedo of 0.04, the detected objects range from 25--44~km in diameter,
and are all well above the survey detection threshold.

No objects are detected beyond 43~AU, though we have the astrometric
sensitivity to detect Solar System members many hundreds of AU
distant, and photometric 
sensitivity to see 37~km objects at 60~AU.  This further tightens the
limits on any ``outer'' Kuiper Belt: any outer component must be
significantly less massive than the known population, or predominately
in the form of objects smaller than 40~km.

In an assemblage of data from ground-based surveys with
well-quantified completeness in magnitude and invariable latitude,
a crude division of the sample into ``classical'' and ``Excited''
dynamical classes reveals a difference in magnitude distribution at
96\% confidence.  Since the two samples have roughly similar distance
distributions, we can infer a true difference in size distribution and
hence in accretion/erosion history.
Both populations are grossly inconsistent with a single-power-law
magnitude or size distribution.
Double power laws give good fits to the present observations, with
rolling power laws slightly less acceptable.
The parametric fits indicate that the Excited population has more of
its mass in large objects, so that the
brightest TNOs are almost entirely in the Excited class, as has been
noted before.  More quantitatively, we find that under the
double-power laws which fit our limited sample, it is feasible
to have a single body as massive as Pluto, {\it i.e.}, Pluto is
uniquely but not anomalously large under this distribution.
This is a sign that the formation of Pluto follows the same physical
mechanism as the other Excited TNOs.  The accretion history of the
Excited bodies placed a larger fraction of the mass into the largest
bodies, whereas the largest CKBO is $\approx60$ times less massive than
Pluto.  This is crudely consistent with scenarios in which the Excited
bodies were formed at smaller heliocentric distances than the CKB
\citep{Ma93,Ma95,Go03}. 

The total mass of the CKB (as we have
defined it, with $i<5\arcdeg$) is constrained to $0.010M_\oplus$, with
accuracy now limited solely by estimates of the albedo, distance, and 
material density of
the bodies.  This is significantly lower than previous estimates, and
exacerbates the difference between the present mass and the $\approx
10 M_\oplus$ initial mass that is believed to be required to
facilitate accretion of the larger TNOs.
The total mass of the Excited sample is less certain because its vertical
and radial extent are poorly known; but the total mass of objects
within 50~AU appears similar to the CKB mass.

The detection of features in the
magnitude (and presumably size) distribution gives fundamentally new
information to be used in constraining the history of the
trans-Neptunian populations, particularly the accretion/erosion
history.  Both dynamical samples show breaks to shallower size
distributions for the $D\lesssim100$~km population. A reevaluation of
the models for collisional evolution is necessary; the assumption of
collisional equilibrium with a scale-free fracturing law may need to
be abandoned.  
Estimates of cratering and resurfacing rates in the trans-Neptunian
region will also require revision.  It is possible, for example, that
collision rates are too low to allow significant exposure of fresh
ices, a mechanism that has been invoked to explain the color diversity
of TNOs \citep{LJ96}.
The CKB seems to be excluded as a source of the JFCs, with the Excited
class being a viable JFC reservoir only with optimistic choices for
model parameters and TNO extrapolations.

The present data are consistent with a picture in which both the CKB
region near 42~AU and a now-vanished inner region of planetesimals
were originally much more massive than at present.  Accretion of 10~km
bodies proceeded in both regions, advancing to 100~km bodies in the
CKB; 1000~km bodies are produced in the inner region, where higher
surface
densities promote more rapid accretion.  The migration of Neptune
(or some other massive body) clears the inner region, scattering a few
percent of the bodies into present ``Excited'' orbits.  This migration
or some other process also excites the eccentricities of the disk,
commencing erosion among TNOs with $D<100$~km.  These smaller bodies
are greatly depleted, today's population being only a fraction of the
fragments produced in early collisions.  Meanwhile, beyond 50~AU there
is either little solid material, accretion never procedes
beyond 10~km bodies, or the modest-sized bodies are destroyed
or removed by subsequent processes.

This scenario has many uncertainties and is hardly unique.  Major open
questions include: in what order and on what time scales did
scattering of the Excited class,
accretion of the largest bodies, and depletion of the small bodies
occur?  Were collisions and depletion important in the last 3 billion years,
with substantial cratering, or has the region been largely dormant?
Was mass loss from the CKB purely due to erosion, or did dynamical
processes eject significant mass in the form of large bodies?  What is
the present source reservoir for
Jupiter-family comets?  Why is there no ``outer'' Kuiper belt?

The quantitative data on TNO size and orbital distributions are
improving rapidly.  In the next year or two, nearly the full available
sky will have been surveyed for $R\lesssim22$ TNOs, using small
telescopes \citep{TB, SEKS}.  Telescopes with effective aperture
$\approx 4$~m have surveyed hundreds of square degrees \citep{DES},
and will survey thousands of square degrees (the CFH Legacy
Survey\footnote{{\tt http://www.cfht.hawaii.edu/Science/CFHLS/}}), or
tens of thousands
(Pan-STARRS\footnote{{\tt http://pan-starrs.ifa.hawaii.edu/}}), in the
coming 5--10 years, to limits of $R\lesssim24$, so a full
characterization of the accretion-dominated size distribution is
forthcoming.  Exploring the erosion-dominated small end is of course
more difficult.  Eight-meter class telescopes can reach $R\approx27$
with great effort, and recovery operations are typically more
expensive than discovery until sky coverage reaches tens of
square degrees.  Wide-area imagers currently coming into operation on
large telescopes will increase the number of known $R>25$ TNO orbits
past the current handful into the tens or hundreds in the next few
years.  Thousands will be possible with a Large Synoptic Survey
Telescope.\footnote{{\tt http://www.lssto.org/}}  The situation for further
explorations at $R>27$ is more bleak: the present study probably
represents the limits of what can be achieved in this decade, unless
the HST is tasked to the problem for $\sim$months of time, or
multi-conjugate adaptive optics becomes capable of providing fields of
view of many arcminutes. A high-throughput space telescope imager such
as SNAP\footnote{{\tt http://snap.lbl.gov}} would be capable of improving on
the current ACS sample by several orders of magnitude circa 2010.
The most valuable information on $D\lesssim50$~km TNOs in this decade
may come from occultation surveys, {\it e.g.}, the nearly operational
TAOS project \citep{TAOS}.
And late next decade, the New Horizons spacecraft will study the
composition and structure of the surfaces of Pluto and one or more
TNOs, providing a cratering and chemical
record.\footnote{{\tt http://pluto.jhuapl.edu/}}
It is already clear that the joint size--dynamical distribution of TNOs
is rich in information to constrain the physical evolution of the
outer Solar System.

\acknowledgements
We thank Tony Roman of STScI for his efforts in successfully
scheduling this intensive program on the HST, and Ron Gilliland for
crosschecks and guidance on the program.  Rahul Dave and Matt Lehner
configured and maintained the computing cluster at Penn.  Jay Anderson
provided distortion maps of the ACS/WFC. Hal Levison provided key
advice based upon experience with the WFPC2 search.  This work
was supported by STScI grants GO-9433.05 and  GO-9433.06. 
Support for program \#GO-9433 was provided by NASA through a grant from
the Space Telescope Science Institute, which is operated by the
Association of Universities for Research in Astronomy, Inc., under
NASA contract NAS 5-26555.
RM
acknowledges support from NASA grants NAG5-10346 and NAG5-11661.

Some of the data presented herein were obtained at the W.M. Keck
Observatory, which is operated as a scientific partnership among the
California Institute of Technology, the University of California and the
National Aeronautics and Space Administration. The Observatory was made
possible by the generous financial support of the W.M. Keck Foundation.

\clearpage

\appendix
\section{Likelihoods for TNO Surveys}
\label{appa}
Correct analyses of the TNO luminosity function $\Sigma(R)$ require a
proper formulation of the likelihood of a given survey detecting a set
of objects at magnitudes $\{m_i\}$.  We cover here some of the
subtleties in this process that we have glossed over in
\S\ref{stats}, and that have been neglected in other works as well.
We then examine how well the simplified analyses described in
\S\ref{stats} approximates the likelihoods which might be
obtained with full attention to the details, and find that, for
current sample sizes, these approximations do not affect the inferred
bounds on the TNO populations.

\subsection{Expected Distribution}
We presume that one is interested in constraining the differential
sky density of sources $\Sigma(R)$ as a function of the magnitude $R$
in some standard system.  Because the detectability may be a function
of some variability parameters $V$, we in fact need to know the joint
distribution $\Sigma(R;V)$ and have
\begin{equation}
\Sigma(R) = \int dV\, \Sigma(R;V).
\label{sigr}
\end{equation}
We assume here that the magnitude $R$ of interest corresponds to the
time average over the light curve.
The sky density will be a function of other variables,
{\it e.g.} dynamical quantities, but we will leave further
generalization to the reader.

The survey may not be conducted in the standard filter band, so we
will denote the mean magnitude in the observed filter band as $\mu$, so
that we have
\begin{equation}
\Sigma_\mu(\mu;V) = \int dR\, P(\mu | R) \Sigma(R;V),
\end{equation}
where we have assumed the object's variability properties to be
independent of color.  We assume
\begin{equation}
\int\! d\mu \, P(\mu|R) = 1.
\end{equation}

A TNO may be detected by the survey, and after detection it will be
assigned an observed magnitude $m$.  The probability
$P(m | \mu;V)$ of being detected {\em and} assigned observed magnitude $m$ is a
characteristic of the survey's methodologies, which can be determined
by proper Monte-Carlo tests.  Note that the detection efficiency for
a chosen TNO magnitude satisfies
\begin{equation}
0 \le \eta(\mu;V) \equiv \int dm \, P(m|\mu;V) \le 1
\end{equation}
because the source may not be detected at all.  We can likewise
integrate over $V$ to define
\begin{equation}
P({\rm detection} | \mu) \equiv \eta(\mu) = { \int dV\, \eta(\mu;V) 
\Sigma_\mu(\mu;V)
	\over
	\int dV\, \Sigma_\mu(\mu;V)}
\end{equation}
or define a detection probability over $R$ as
\begin{equation}
\eta(R) = { \int d\mu\, dV\, \eta(\mu;V) P(\mu|R) \Sigma(R;V)
	\over
	\int dV \,  \Sigma(R;V)}.
\label{etar}
\end{equation}
Note that we can determine $\eta(\mu;V)$ strictly from analysis of the
survey characteristics, because $\mu$ and $V$ fully specify the behavior
of objects in the imaging.  But to estimate $\eta(\mu)$ or $\eta(R)$, we
also need the distributions of the TNO population over variability
and/or color, unless the detection efficiency is sufficienctly
independent of them.

The expected distribution of detections from a survey of solid angle
$\Omega$ is
\begin{equation}
g(m) \equiv {dN \over dm} 
= \Omega \,\int d\mu\,dV\, \, P(m | \mu;V) \int dR \, P(\mu|R) \Sigma(R;V).
\end{equation}
In simple words, the observed distribution $g(m)$ is the convolution
of the input distribution $\Sigma(R)$ with the magnitude shifts
induced by color, variability, measurement error, and detection
efficiency.   Many real surveys are inhomogeneous, with $P(m | \mu;V)$
varying with location or time as weather conditions or integration
times vary, and in these cases the expected $g(m)$ must further be
convolved over these variable conditions.

\subsection{Likelihoods}
Given the function $g(m)$ derived from some $\Sigma(R,V)$, we can
assign a likelihood to detection of
a set $\{m_1, m_2, \ldots, m_M\}$ of detections in the survey to be
\begin{equation}
L(\{m_i\}|\Sigma) = dm^M \exp \left[-\int dm\, g(m)\right] \, \prod_{i=1}^M
g(m_i).
\label{likeg}
\end{equation}
The form of the likelihood is easily derived by considering each
interval $\delta m$ of magnitude to be a Poisson process with probability
$\delta P = g(m)\,\delta m$ of having one detection and $1-\delta P$ of zero
detections.

\subsubsection{Comparison to Gladman \etal\ Form}
\citet{Gl98} provide an alternative form of this equation which in
our notation and suppressing the variability parameters, is
\begin{equation}
L(\{m_i\}|\Sigma) = \exp \left[-\Omega \int dR\, \eta(R) \Sigma(R)\right] \, 
	\prod_{i=1}^M \int dR \, \ell_i(R) \Sigma(R).
\label{gll}
\end{equation}
The function $\ell_i(R|m)$ ``describes the uncertainty for the [true]
magnitude of object $i$'' according to \citet{Gl98}, but the exact
meaning of this is not well understood in the community.  It does {\em
not} mean a simple convolution with a Gaussian error term centered on
$m_i$. 

The identification of \eqq{gll} with \eqq{likeg}, up to factors of
$dm$ and $\Omega$ that are independent of the $\Sigma$ model, is clear from
the following.  The exponentiated quantity is in either case $\bar N$,
the expected number of detections, given by
\begin{eqnarray}
\label{nbar}
\bar N \equiv \int dm\, g(m) & = &
	\Omega \int dm\, \int d\mu\,dV\, \, P(m | \mu;V) \int dR\, P(\mu|R)
	\Sigma(R;V) \\
 & = & \Omega \int dR \int dV\,dm\,d\mu\,  P(m | \mu;V) P(\mu|R) \Sigma(R;V)
	\\
 & = & \Omega \int dR\, \eta(R) \Sigma(R),
\end{eqnarray}
making use of (\ref{sigr}) and (\ref{etar}).  

To equate the terms under the product sign, we note that, if the
detection probability is independent of variability, we can write
\begin{equation}
g(m) = \Omega \int \! dR\, P(m|R;V) \Sigma(R),
\end{equation}
so that if we identify $\ell_i(R)=P(m_i | R;V)$, then \eqq{gll}
becomes equivalent to (\ref{likeg}).  Recall that $P(m|R;V)$ is the
probability of the TNO at $R$ being {\em detected} at $m$, so its
integral over $m$ is $\eta(R)\le1$, and there is in any case no need to
have $\int dR\, \ell_i(R)=1$.  It is hence not appropriate to take
$\ell_i(R)$ to be a Gaussian about $m_i$ with dispersion $\sigma$
given by a rough magnitude uncertainty.  

A simplification is possible in the case $P(m |
R;V)=\eta(R)P(m-R)$, meaning that
probability of detection $\eta(R)$
is independent of the observed magnitude $m$ for given $R$, and the
magnitude measurement error has a fixed distribution with unit
normalization.  This would
be the case when the true magnitude is measured in followup
observations that are distinct from the discovery observations.  In
this case we have
\begin{eqnarray}
g(m) & = &  \int \!dR\, P(m-R) [\Omega\eta(R)] \Sigma(R) \\
\Rightarrow \ell_i(R) & = & P(m-R) \Omega_{\rm eff}(R).
\end{eqnarray}
The \citet{Gl98} likelihood can be misinterpreted as assuming that
$\ell_i(R)=P(m_i-R)$.  We see that this is incorrect---even in this
simplified case it is necessary to have an additional term of
$\eta(R)$ or $\Omega_{\rm eff}(R)$ under the $R$ integral.
In many surveys, however, the detection probability is highly
correlated with the observed magnitude $m$, so even this simple form 
is not applicable.

\subsubsection{Fitting the Amplitude}
\label{marginsection}
As a useful aside, consider the maximum-likelihood estimate of the
scale factor $s$ that normalizes a candidate TNO distribution function
of the form
\begin{equation}
\label{scalefactor}
\Sigma(R;V) = s \Sigma_o(R;V).
\end{equation}
The expected distribution will clearly scale as $g(m)=sg_o(m)$, with
the obvious notation.
The likelihood (\ref{likeg}) then has all of its dependence on $s$ in
terms
\begin{equation}
L \propto s^M e^{-sN_o},
\end{equation}
where $N_o$ is the expected number of detections $\bar N$ from
\eqq{nbar} for $\Sigma=\Sigma_o$.  
If we adopt a logarithmic
prior $p(s)\propto s^{-1}$, then the Bayesian expectation values and
probability distribution for $s$ are
\begin{eqnarray}
\label{logprior}
\bar s \equiv \langle s \rangle & = & M/N_o \\
\langle N \rangle & = & M \\
p(s>s_o) & = & P(M,s_oN_o),
\end{eqnarray}
where the last right-hand side is the incomplete gamma
function.\footnote{For $M=0$ (no detections), the logarithmic prior
leads to unnormalizable posterior distributions, and we revert to a uniform prior.}

If we are interested only in the shape $\Sigma_o$ of the TNO
distribution, we will marginalize over the nuisance parameter $s$,
which for the logarithmic prior yields
\begin{eqnarray}
L(\{m_i\} | \Sigma_o) & = & \int ds\, p(s) L(\{m_i\} | \Sigma_o, s)
\\
 & = & N_o^{-M} \Gamma(M) \prod_i g_o(m_i) \\
\label{smargin}
 & = & M^{-M} \Gamma(M) \prod_i \bar s\, g_o(m_i) \\
 & = & M^{-M} \Gamma(M) e^M L(\{m_i\} | \Sigma_o, \bar s).
\end{eqnarray}
This analytic marginalization speeds up our likelihood analyses for
the shape of $\Sigma$.  Note that the formulae fail for $M=0$, which
is not surprising, since we cannot constrain the shape of $\Sigma$ if
we have no detections.

\subsection{Detection Cutoff}
In the following we derive some exact results for
simple cases, and then examine the error made in simple approximations.
We will assume that in the neighborhood of the survey limit, the source
density behaves as $\Sigma(R)=\Sigma_o 10^{\alpha R}$.

The simplest case to analyze is when the survey assigns a detection
magnitude $\rho_i$ to each TNO, and those brighter than a cutoff $\rho_o$
are accepted into the survey.  From measurement noise alone,
the
detection magnitude usually has a Gaussian distribution about the true
magnitude $R$ with a standard deviation of $\sigma(R)$ mag.  If color
and variability are important, we can consider these as additional
sources of scatter in the $P(\rho | R)$ distribution, as quantified in
later sections, for now we consider measurement noise alone.

Consider two cases:  in the first, which we'll call a {\em snapshot}
survey, the survey is halted after its
detection observations, so that $m_i=\rho_i$.   If we further assume
that $\sigma(R)$ varies little near the completeness limit, then we have
\begin{eqnarray}
\label{gocutoff}
g(m) & = & \Omega\Sigma_o 10^{\alpha m} 
\exp[\alpha^2 \ln^2(10)\,\sigma^2/2] \,\Theta(\rho_o-m) \\
\bar N & = & { \Omega\Sigma_o \over \alpha \ln 10}
 10^{\alpha \rho_o} \exp[\alpha^2 \ln^2(10)\,\sigma^2/2] \\
\eta(R) & = & (2\pi\sigma^2)^{-1/2}
\int_{-\infty}^{\rho_o} d\rho \, \exp[-(R-\rho)^2/2\sigma^2] \\
 & = & 
\frac{1}{2} \, {\rm erfc}[(R-\rho_o)/\sqrt{2}\sigma].
\label{erfeta}
\end{eqnarray}
Here $\Theta$ is the step function.
In this case, ignoring the difference between observed and detected
magnitudes---that is, using $\Omega\Sigma(R)$ instead of $g(m)$ in the
likelihood (\ref{likeg})
for the $\{m_i\}$---amounts to a normalization error of 
$\exp[\alpha^2 \ln^2(10)\,\sigma^2/2]$.
The completeness follows \eqq{aeff} with $w=\sqrt{2}\sigma$.  

In a second type of survey, which we'll call a {\em magnitude
followup} survey, the
detected objects are followed up with more integration, so that the
errors on magnitude are substantially reduced, and we end up with
$m_i=R_i$, in the absence of color terms. Our ACS survey is well
approximated in this way, since the
faintest detected object has $S/N>20$ in the full dataset, hence a
magnitude uncertainty (in F606W) of $<0.05$~mag.
In this case we have
\begin{equation}
g(m)  =  \Omega \Sigma(R) \eta(R),
\end{equation}
where the detection efficiency $\eta(R)$ is from \eqq{erfeta},
and is in fact independent of the form of $\Sigma(R)$. 
We note that in the case of a magnitude-followup survey, the
approximate likelihood in \eqq{lapprox} is exactly correct.

\subsubsection{Size of the Errors}
Suppose we were to conduct a snapshot survey that detects objects at
$\{m_i\}$, and also a Monte Carlo survey to determine the form of
$\eta(R)$, but then we get lazy in implementing \eqq{likeg} and ignore
the difference between $m$ and $R$ in calculating our likelihoods, using
\begin{equation}
g(m) = \Omega\Sigma(m)\Theta(\mu_o-m)
\label{gapprox1}
\end{equation}
instead of the proper form in \eqq{gocutoff}.
Then we see that we are in effect using a luminosity
function that is in error by the normalization factor
$F=\exp[\alpha^2 \ln^2(10)\,\sigma^2/2]$.
We can bound this factor by noting that
a TNO detection must have $S/N>6$ to be reliably distinguished
from noise in most detection schemes, so we expect the noise on the
detected magnitude to be $\sigma<2.5/[\ln(10)\,(S/N)]\le 0.18$~mag.  In
this case we have $F \approx 1+0.04(\alpha/0.7)^2$.  
First we note that this bias factor, if constant across all survey
data, only affects the normalization of $\Sigma$ and does not affect
our conclusions about the shape of $\Sigma$.  For $|\alpha|\le0.7$,
the resultant 4\% bias would be smaller than the Poisson errors in 
our analysis, which contains only 130
total objects and is hence incapable of detecting $<10\%$ shifts in
the presumed $\Sigma(R)$ at any magnitude.

If the bias factor $F$ varies with $R$ than we could infer an
incorrect shape for $\Sigma$, but again with our sample of 130~TNOs
the difference is below the Poisson noise.  Note that we conclude
$|\alpha|\lesssim0.7$ for all samples and magnitudes, save perhaps the
bright end of the CKBO population, where we could incur a bias
somewhat above 10\%---but the Poisson noise is also higher in this
subsample, so again the bias is negligible.
Note that for surveys with significant magnitude
followup, there are no 
approximations in taking $g(m)=\Omega\eta(m)\Sigma(m)$, so there is no bias
at all.  This is the case for our ACS data, and also for the La and TB
data, since we have accurate followup mean $R$ magnitudes for nearly all
detections. 

\subsubsection{Inhomogeneous Surveys}
If we take this simple case literally, we would expect all surveys to
report a completeness function $\eta(R)$ that follows \eqq{aeff} with
$w\approx
(S/N)^{-1}/\sqrt{2}$.  This is true in our case, with a limiting
$S/N=8.2$ and $w=0.08$.
But most surveys report completeness functions $\eta(R)$ that are
significantly broader than this form.  This is generally due to
inhomogeneity, as ground-based surveys must deal with varying seeing
and coverage, etc., that the ACS survey does not.   In effect the
survey area $\Omega$ is divided into sub-areas $\Omega_k$ that have cutoffs
$\rho_k$.  The expected distributions $g_k(m)$ for each sub-survey may
be summed to yield the total $g(m)$ for the survey.  It is convenient
to define the total survey area which reaches a chosen threshold:
\begin{equation}
\Omega_{\rho}(\rho) \equiv \sum_{\rho_k>\rho} \Omega_k.
\end{equation}

For a {\em snapshot} survey, we obtain:
\begin{eqnarray}
\label{ihcutoff}
g(m) & = & F \Omega_\rho(m) \Sigma_o 10^{\alpha m} \\
\label{iheta}
\eta(R) & = & (2\pi\sigma^2)^{-1/2}
\int d\rho \, 
{\Omega_{\rho}(\rho) \over \Omega} \exp[-(R-\rho)^2/2\sigma^2] 
\\
 & = & 
\frac{1}{2A} \int d\rho {d\Omega_\rho \over d\rho} \,{\rm erfc}[(R-\rho)/\sqrt{2}\sigma].
\end{eqnarray}
The falloff in completeness $\eta(R)$ is now the convolution of the
inhomegeneity with the measurement error, so will be more extended
than either.  But the bias $F$ is {\em not} affected by the
inhomogeneity, only by the measurement error, so we see that the
approximation in \eqq{gapprox1} can be replaced with
\begin{equation}
\label{gapprox2}
g(m)\approx \Omega_\rho(m)\Sigma(m)
\end{equation}
with equal accuracy, sufficient for our purposes.

For an inhomogeneous {\em magnitude-followup} survey, we find that
once again \eqq{lapprox} is exact.

\subsection{Color Errors}
The comparison of the observed-band magnitude $\mu$ with a density function
$\Sigma$ defined over a standard magnitude $R$ can be thought of as a
convolution of the expected $\Sigma(R)$ with the color distribution
$P(\mu|R)$.  We have assumed a single value for our $F606W-R$ color
conversion, but there is a spread in colors which may be viewed as an
additional source of stochastic uncertainty on the $R$ magnitude of
our detections.  The standard deviation of $V-R$ colors is 0.10~mag
\citep{TR03, TRC03}.   Since $F606W$ is nearly the union of $V$ and
$R$, similar to the ``$V\!R$'' filter used by several ground-based
surveys, we estimate the scatter in $F606W-R$ to be $\sigma_{V\!R}\approx
0.05$~mag. 

This scatter may be treated as a measurement error on $R$ that is
added in quadrature to the measurement noise.
The $<0.1$~mag uncertainty
due to the color conversion should have negligible effect on the
interpretation of $\Sigma(R)$, at least for the surveys under current
consideration which have $<100$ detections and are in well-calibrated
filters bounded by the $V$ and $R$ passbands.  Note, however, that the
two brightest surveys, TB and La, have very poorly defined filter
bands, so the color uncertainties could be larger here.  For most of these
bright objects, however, a high-precision $R$ magnitude has been found
in the literature, so the color or photometry uncertainties are minimal.

\subsection{Variability}
The variability of the TNOs can influence both the detectability and
the observed magnitude $m_i$ that is assigned to a discovery.
The effect of variability on the detection function $P(m|\mu;V)$ has not
been accurately quantified for any published survey, and furthermore
the variability dependence of $\Sigma(R,V)$ is very poorly
constrained, so at this point it is basically impossible to
meaningfully incorporate variability into the likelihood functions.

We can nonetheless show with some basic modelling that the effect of
variability on current analyses should be minimal given the present
Poission uncertainties.  The true form of $P(m|\mu;V)$ for real-life
surveys is probably rather complex:
most surveys require detection in two or
more observation sets seperated by hours to days, and the known
variable TNOs have periods of 3--13 hours.
We will consider some simplified forms of variability selection.

\subsubsection{Snapshot Survey}
A ``snapshot'' survey in this context observes the TNO only for a time
short compared to 
the lightcurve period, so the observed magnitude $m_i$ is an
instantaneous sample of the light curve, and the detection probability
is a function solely of $m_i$.  We ignore color effects and
measurement noise here, for simplicity. 
For the snapshot survey, let $R(t)$ be the magnitude at time $t$.  We
then have
\begin{equation}
P(m | R;V) = \Theta(m-\rho_o) P(m | R(t)) P(R(t) | R;V),
\end{equation}
where $\Theta$ is a step function. $P(R(t)|R;V)$ is simply the
distribution of instantaneous magnitudes for the chosen light curve.
So $g(m)$ is the convolution of the intrinsic
distribution $\Sigma(R)$ with the distributions due to
variability and measurement noise.  The variability in this case just
acts to broaden the detection cutoff vs mean magnitude, just like a
source of measurement noise.

How broad is the variability kernel?
For a sinusoidal light curve with peak-to-peak amplitude $A$, the
distribution $P(m-R)\propto [1-(m-R)^2/4A^2]^{-1/2}$.  When a
power-law $\Sigma(R)$ is convolved with this distribution, the
resultant $\Sigma_m\equiv dN/dm$ is enhanced by the factor
\begin{equation}
\label{varmag}
F = 1 + (A \alpha \ln 10)^2/16 + (A \alpha \ln 10)^4/1024 + \ldots
\end{equation}
Even in the worst case of a square-wave light curve, the bias factor
is $F=\cosh(A\alpha \ln(10)/2)$, which is $\approx 2$ times larger.

Our \newbf, along with 2001~QG$_{298}$, are the most highly
variable of the $\approx60$ TNOs tested for variability to date, with
$A\approx1.1$~mag \citep{lcpaper, SJ04}.  
With this $A$ and $\alpha\le 0.7$, the surface density for
instantaneous magnitudes is $F\le 1.20$ higher than the $\Sigma(R)$
for mean magnitudes. 

One might expect small (faint) TNOs to be less symmetric and
hence more variable than bright TNOs, though this is not yet borne
out.  But it can never be the case that
they have uniformly high amplitudes $A$, because pole-on rotators
must have zero amplitude.  The formulae of \citet{LL03} for the
light-curve amplitudes of triaxial ellipsoids in minimum-energy
rotation can be used to show that a population of identically-shaped
TNOs with randomly oriented rotation axes will have a roughly uniform
distribution of light-curve amplitudes between 0 and $A_{\rm max}$.
We might expect a similar result for variability due to surface
features rather than gross body shape.  In this case the
population-averaged enhancement factor will be
\begin{equation}
\label{fmean}
F \approx 1 + (A_{\rm max} \alpha \ln 10)^2/48.
\end{equation}
Take, as a pessimistic case, a population of TNOs that are all
sufficiently asymmetric to have light-curve amplitudes as large as
\newbf\ if viewed 90\arcdeg\ from the rotation pole.  Then we find the
difference between the instantaneous- and mean-magnitude distributions
to be $F\le1.07$, less than 7\%.  Thus the error made by failing to
distinguish these cases is below the Poisson noise of the current
samples.

\subsubsection{Worst-Case Models}
The worst possible bias in the TNO sample due to variability would be
if the selection were based on peak (or minimum) flux over the
light curve.  A simplistic scenario that might spawn such a selection
effect is for a survey that 
consists of many repeated observations of
each target, and a ``detection'' consists of the TNO rising above the
threshold in {\em one or more} of the epochs.  Once detected, the TNO is
found in all epochs, and we report a mean magnitude.  In this
scenario, the reported magnitude is an unbiased estimator of the true
$R$, but the detection criterion is a step function of the peak
flux over the light curve, $\eta(R,V)=\Theta[\rho_c-(R-A/2)]$, for a
light curve with peak-to-peak amplitude $A$.
Note that in this case, as for the
snapshot model in the previous paragraph, there is a bias
{\em toward} detection of highly variable TNOs.  Hence increased
variability in the faint population would tend to {\em steepen} the
derived $\Sigma(R)$, as opposed to the flattening we detect in the ACS
sample. 

Conversely imagine that the detection criterion is that the TNO be
above threshold in {\em all} of the epochs.  In this case $\eta(R,V)$
will be a function solely of the {\em minimum} flux over the
light curve, and the detection is biased {\em against} high
variability at a given mean $R$.

Again take a pessimistic view that $A$ is uniformly distributed
between 0 and $A_{\rm max}\approx1$~mag.  Then (in the absence of
noise) the selection function $\eta(R)$ drops linearly from unity at
$\rho_o-A/2$ to zero at $\rho_o$.  This closely resembles the
error-function form in \eqq{aeff} with $m_{50}=\rho_0-A_{\rm max}/4$ and
$w=A_{\rm max}/8$.  Hence in this worst-case scenario, the effect of
variability is to shift our completeness limit brightward by
$\le0.3$~mag and broaden the cutoff by $\le 0.15$~mag.

\subsection{Is a Simple Treatment Sufficient?}
In analysing the surveys taken to date, we have taken a simplified
form for the likelihood, embodied by \eqq{lapprox}.  We have seen that
this form is exact in the case of a magnitude-followup survey with a
fully realistic Monte-Carlo derivation of $\Omega_{\rm eff}(R)=\Omega\mu(R)$
that is properly marginalized over the variability and color of the TNOs.
Technically the simplified likelihood is incorrect for our analysis
because: 
\begin{itemize}
\item The middle-magnitude surveys have no followup and are
``snapshot'' surveys, meaning that $g(m)$ from \eqq{ihcutoff} is more
appropriate.  We have shown that the $F$ factor is smaller than
Poisson errors, and the difference between $\Omega_\rho(m)$ and $\eta(m)$
is only a convolution by the relatively sharp function \eqq{erfeta}
from magnitude errors of $\sigma\lesssim0.2$~mag.  If we avoid the
regions where $\eta<0.2$, then the effect on the likelihood is small.
\item The Monte-Carlo estimates of $\eta(R)$ in most surveys ignore
errors in color terms.  The effect is to broaden $\eta(R)$ by
$\lesssim0.05$~mag in quadrature.
\item The Monte-Carlo estimates of completion ignore variability.  For
short-exposure surveys this is equivalent to a $<7\%$ mis-measure of
the underlying $\Sigma(R)$.  In the worst imaginable case of selection
effects, it could lead to a shift in the completeness limit by
$\le0.3$~mag and a broadening of $\le0.15$~mag.
\item For technical reasons the Monte-Carlo for the ACS data yields
$\Omega_\rho(m)$ rather than the slightly broader $\eta(R)$ function.
\item Some readers may not trust the completeness tests or magnitude
error estimates of our or other surveys.
\end{itemize}
To test the effect of these approximations on our results, we have
re-analysed the survey data after artificially broadening the $\eta(R)$
cutoff width $w$ by 0.3~mag for the ACS and/or TB surveys.  We also
shift the cutoff $m_{50}$ of the ACS completion function brightward by
0.3~mag to test the import of a substantial mis-estimate of our
completeness or variability treatments.

Note that we do not apply these completeness changes to all the
surveys, just the ones at the ends of the magnitude ranges, because we
are more interested in errors that might have influenced the shape of
the $\Sigma(R)$ fits.  We find that no combination of these tweaks to
the assumed $\eta(R)$ functions causes a significant shift in the
$\{\alpha_1, \alpha_2\}$ contours of Figure~\ref{a1a2contour} or in
any of the quantities calculated herein. 
When we shift $m_{50}$ of the ACS survey brightward, the ability to
discriminate $\alpha_2$ for the two classes is weakened 
(distinct $\alpha_2$ values are still strongly suggested), and the
extrapolation to the JFC source population can be higher.
Our main conclusions, however, remain
valid: a break in the power-law $\Sigma(m)$ is required, there is a
significant deficit of small bodies, the CKB and Excited classes have
distinct size distributions, with the latter having clearly shallower
$\alpha_1$. 
We therefore conclude that a
simplified treatment of the likelihoods is currently adequate.  Simply
put, the ACS survey has only 3 detections, so there is no need to be
too careful with subtle variability or color effects.

When uniform TNO surveys with $\gg100$ detections are available,
however, a more careful treatment will be required.  In particular,
careful attention to the effect of variability may be needed, and the
likelihood analyses should perform the proper integrals to get $g(m)$
in the case that accurate followup magnitudes are not available,
because the distinction between the discovered magnitude $m$ and the
true magnitude $R$ is important at the few-percent level.

\end{document}